\documentclass[letterpaper,twocolumn,10pt]{article}
\usepackage{nsdi}

\usepackage{tikz}
\usepackage{amsmath}

\usepackage{filecontents}

\usepackage[english]{babel}
\usepackage{blindtext}

\usepackage{tikz}
\usepackage{amsmath}
\usepackage{amsfonts}

\usepackage{hyperref}
\usepackage{url}
\usepackage{graphicx}
\usepackage{wrapfig}
\usepackage{tabularx}
\usepackage{color, colortbl}
\usepackage{physics}
\usepackage{booktabs}
\usepackage{multirow}
\usepackage{xcolor}
\usepackage{graphicx}
\usepackage{footnote}
\usepackage{tabularx}
\usepackage{float}
\usepackage{lipsum}
\usepackage{multicol}
\usepackage{xspace}
\usepackage{siunitx}
\usepackage{enumitem}
\usepackage[font={small,bf}]{caption}

\newcommand{\namereparo} {Reparo\xspace}
	
\definecolor{LightCyan}{rgb}{0.88,1,1}
\definecolor{Grey}{rgb}{0.93,0.93,0.93}
\definecolor{DarkGrey}{rgb}{0.55,0.55,0.55}

\newcommand{\red}[1]{\textcolor{black}{#1}}

\newcommand{\Fig}[1]{Fig.~\ref{fig:#1}}
\newcommand{\Sec}[1]{\S\ref{sec:#1}}

\newcommand{\Tab}[1]{Tab.~\ref{tab:#1}\xspace}
\newcommand{\NewPara}[1]{\noindent{\bf #1}}

\colorlet{darkgreen}{green!65!black}
\colorlet{darkblue}{blue!75!black}
\colorlet{darkred}{red!80!black}
\definecolor{lightblue}{HTML}{0071bc}
\definecolor{lightgreen}{HTML}{39b54a}
\definecolor{shadecolor}{RGB}{150,150,150}
\newcommand{\magecolor}[1]{\par\noindent\colorbox{shadecolor}}

{
\setlength{\leftmargini}{9pt}%
\begin{itemize}%
\setlength{\itemsep}{0pt}%
\setlength{\topsep}{0pt}%
\setlength{\partopsep}{0pt}%
\setlength{\parskip}{0pt}}%
{\end{itemize}}

\graphicspath{{./figures/reparo}}

\begin{document}
\title{\namereparo: Loss-Resilient Generative Codec for Video Conferencing}

\author{\textnormal{Tianhong Li
\quad Vibhaalakshmi Sivaraman 
\quad Pantea Karimi
\quad Lijie Fan} \\
\textnormal{
Mohammad Alizadeh 
\quad Dina Katabi} \\
\quad \\
MIT CSAIL \\
\quad \\
}


\maketitle
\begin{abstract}

Packet loss during video conferencing often results in poor quality and video freezing. Retransmitting lost packets is often impractical due to the need for real-time playback, and using Forward Error Correction (FEC) for packet recovery is challenging due to the unpredictable and bursty nature of Internet losses. Excessive redundancy leads to inefficiency and wasted bandwidth, while insufficient redundancy results in undecodable frames, causing video freezes and quality degradation in subsequent frames.

We introduce Reparo — a loss-resilient video conferencing framework based on generative deep learning models to address these issues. Our approach generates missing information when a frame or part of a frame is lost. This generation is conditioned on the data received thus far, considering the model's understanding of how people and objects appear and interact within the visual realm. Experimental results, using publicly available video conferencing datasets, demonstrate that Reparo outperforms state-of-the-art FEC-based video conferencing solutions in terms of both video quality (measured through PSNR, SSIM, and LPIPS) and the occurrence of video freezes.

\end{abstract}

\section{Introduction}

Video conferencing applications are a crucial part
of modern life. Despite all the advancements, video conferencing applications still suffer from packet loss, resulting in diminished quality and video freezing~\cite{boyce1999packet,zheng2001improved,tambur}. This problem is exacerbated by the strong dependence between encoded frames in traditional video codecs~\cite{vp8,vp9,h264,h265,av1}. For instance, in traditional codecs, P-frames (Predicted picture) depend on previous I-frames (Intra-coded picture); hence, the loss of an I-frame affects many subsequent frames, causing a jarring experience where the video freezes and subsequently exhibits poor quality until the codec can recover from these dependencies.

Existing systems employ two main techniques to combat this problem: retransmission and forward error correction (FEC). Since real-time applications such as video conferencing must recover lost packets within a limited latency to meet the real-time playback requirement, retransmission is only suitable for scenarios with short round trip times. In all other cases, such applications rely on FEC to recover lost packets within acceptable latency. FEC schemes send redundant packets, known as “parity” packets, to recover the lost data using traditional block codes \cite{reed1960polynomial, mackay2005fountain} or latency-optimized streaming codes \cite{tambur}. However, all FEC approaches face a challenge in choosing how much redundancy to add since losses in the Internet are bursty and unpredictable. Too much redundancy leads to inefficiency and wasted bandwidth, and too little redundancy leads to undecodable frames, causing video freezes and quality degradation in subsequent frames.

This paper presents a novel approach to loss recovery in video conferencing, without using redundant packets or retransmission requests. Instead, when a loss occurs, the receiver leverages the power of "generative" models to reconstruct the missing information. This approach builds upon recent advancements in generative deep learning models \cite{li2022mage, li2023self, liu2023audioldm, rombach2022high, MAE}, which are capable of reconstructing images (i.e., frames) even when a significant portion of the data is missing \cite{MAE, li2022mage}. Unlike traditional video codecs, which solely rely on received data for frame reconstruction, generative models operate similarly to humans. They utilize a wealth of knowledge about how people and objects appear, move, and interact to generate the missing information. For instance, when presented with one eye of a person, they can generate the missing eye, or with a partial view of an arm, they can reconstruct the entire torso. The key insight here is that such a generative model is ideally suited for a loss-resilient video codec. However, instead of being guided by textual prompts as seen in typical image generators like DALL-E 2, it can be directed by the received pixels in the current frame, as well as information from past frames, to generate the missing content. This guidance ensures that the generated pixels remain compatible with the correctly received data in the current frame and previous frames, eliminating the potential for hallucinating incompatible content, while generating the data in the lost packets. 

\begin{figure}[t]
\begin{center}

\includegraphics[width=1.0\linewidth]{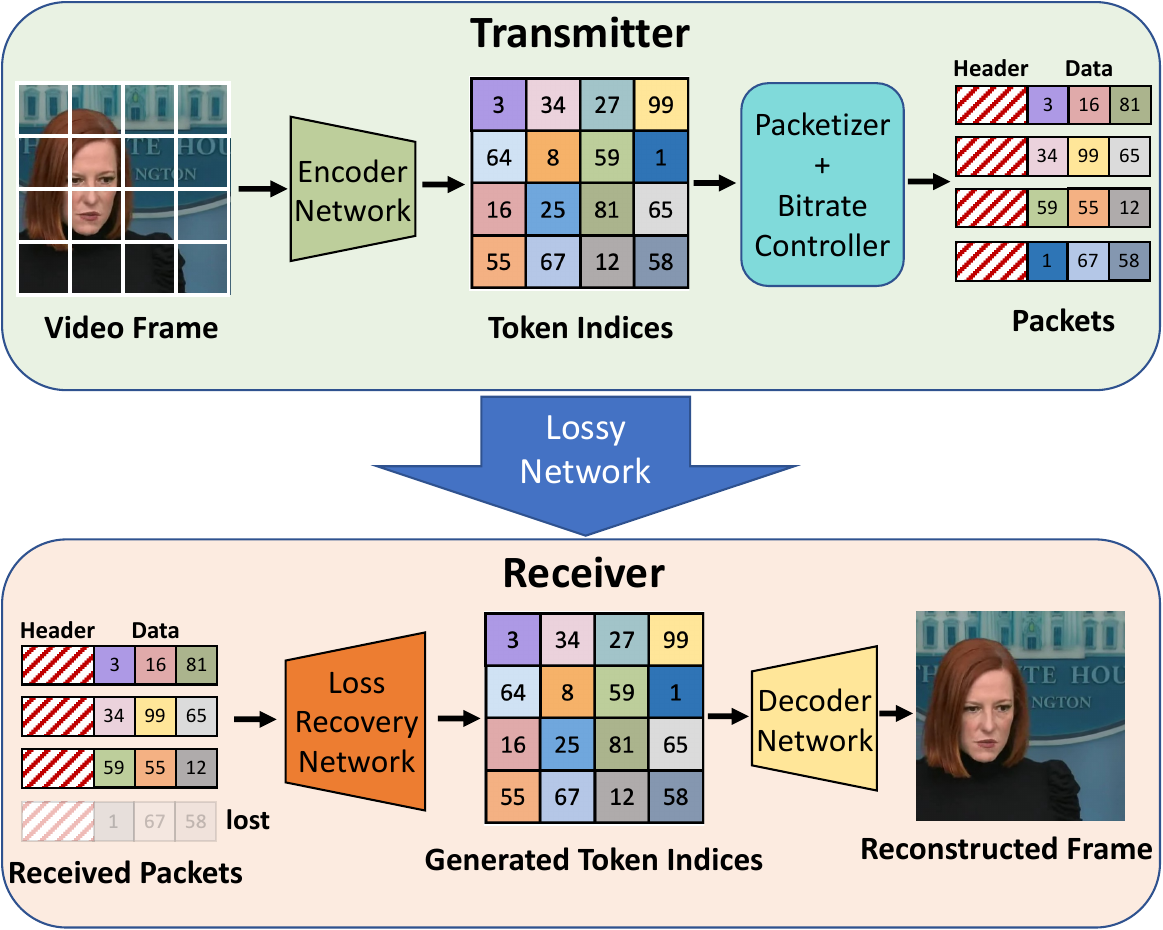}
\end{center}
\caption{Overview of \namereparo. It comprises an encoder-decoder pair responsible for converting RGB frames into quantized tokens and vice versa, as well as new modules for packetization, bitrate control, and loss recovery that operate in the token space.}\label{fig:overview}
\vspace{-10pt}
\end{figure}

We introduce \namereparo, a loss-resilient generative codec for video conferencing. The design of \namereparo involves two steps. In the initial step, it learns to represent the specific video domain of interest, namely video conferencing, using a small codebook of visual tokens, where each token refers to a patch in a frame. \namereparo then operates on these tokens. The transmitter uses a neural network to encode each frame into its respective set of tokens, packetizes them, and transmits the data. Some of these packets may get lost in the network. The \namereparo receiver has a neural network that can regenerate the missing tokens from those it receives and its knowledge of how tokens relate to each other in the visual world. Finally, the reconstructed tokens are decoded to produce the original frame. \Fig{overview} illustrates the components of \namereparo.

In addition to its resilience against packet loss, \namereparo offers three notable advantages:

\begin{enumerate}[noitemsep,topsep=0pt,parsep=0pt,partopsep=0pt]

\item \textbf{Efficient Compression}: \namereparo efficiently compresses data by capturing prevalent visual features and dependencies among objects and shapes within its codebook. This codebook is pre-learned and known to both the transmitter and receiver, allowing \namereparo to transmit only token indices instead of the actual tokens and their underlying dependencies.

\item \textbf{Target Bitrate Compliance}: Traditional video codecs exhibit variable bitrates partially due to the significant size discrepancy between I-frames and other frames. This inconsistency makes it challenging for these codecs to meet a precise target bitrate, leading to fluctuations during transmission, transient congestion, and an elevated risk of packet loss or delay. \namereparo, in contrast, maintains a constant bitrate as all frames are treated equally, making it easy to adapt to any desired bitrate.

\item \textbf{One-Way Communication}: Video conferencing frameworks, including those that rely on FEC \cite{tambur, webrtc}, typically rely on the receiver to send an ACK for every decodable frame. The sender waits for the ACK to retransmit, which could cause even longer freezes when the round-trip time is long. In contrast, in \namereparo, the receiver does not need to communicate with the transmitter about undecodable frames; it will always use the received tokens to reconstruct the lost ones. 

\end{enumerate}

We have conducted an extensive evaluation of \namereparo, comparing it with FEC schemes integrated with WebRTC \cite{webrtc} (ULPFEC and flexFEC), and Tambur \cite{tambur}, a state-of-the-art streaming-code-based FEC approach. The evaluation uses a large corpus of publicly available video clips spanning 5 hours from 84 individuals, which is significantly larger and more diverse than the validation set in prior works on video conferencing \cite{sivaraman2022gemino, tambur, cheng2023grace}. The results are as follows:

\begin{enumerate}[noitemsep,topsep=0pt,parsep=0pt,partopsep=0pt]
\item \namereparo consistently improves the visual quality of displayed videos across all loss levels. It achieves 33.4 dB, 32.9 dB, and 31.6 dB for the 10\% worst PSNR (Peak Signal to Noise Ratio) values under low, medium, and high loss levels, respectively, outperforming state-of-the-art integration of VP9 (a classical video codec)+Tambur by 11.5 dB, 16.4 dB, and 14.7 dB. Notably, this loss resilience does not come at the expense of efficient coding; \namereparo achieves similar or better PSNR compared to baselines in the absence of packet loss.   

\item \namereparo\ nearly eliminates video freezes and significantly reduces the number of unrendered frames compared to the baselines. Under low, medium, and high loss rates, \namereparo fails to render only 0.2\%, 0.8\%, and 2.0\% of frames, while VP9+Tambur fails to render 8.0\%, 13.1\%, and 29.2\% of frames, respectively.

\item In rate-limited environments, \namereparo optimally utilizes the full link capacity by consistently transmitting at a fixed desired bitrate. In contrast, VP9+Tambur must maintain a lower average bitrate to prevent packet loss due to the VP9 encoder's bitrate variability. This results in a higher PSNR for \namereparo compared to VP9+Tambur (35 dB vs. 33.4 dB).

\end{enumerate}

Our \namereparo codec implementation runs in realtime on a V100 GPU, at the transmitter and the receiver, which is comparable to the Apple M2 Max GPU in Macbook Pro laptops. Computational requirements are expected to improve over time with the integration of more powerful GPUs into consumer devices.

\red{In summary, video communication has involved a trade-off between efficiency and resilience traditionally. To maximize efficiency, codecs encode frames together (as a delta from a reference frame), whereas to maximize resilience, each frame should be encoded separately. \namereparo stands out as the first codec to encode each frame \emph{independently}, with no reliance on other frames, while maintaining efficiency akin to state-of-the-art video conferencing codecs that encode frames together. We believe that \namereparo underscores the potential of interdisciplinary design, marrying advances in computer vision with core principles in coding theory and communication systems.}

\section{Related Work}
\label{sec:related}


\NewPara{Video Codecs.}
Video applications typically use classical codecs such as VP8, VP9, H.264, H.265, and AV1 ~\cite{vp8,vp9,h264,h265,av1}. These codecs compress video frames using block-based motion prediction, separating them into keyframes (I-frames) that are compressed independently and predicted frames (P-/B-frames) that are compressed based on differences between adjacent frames. While classical codecs are widely supported and efficient in slow modes, in real-time video conferencing modes, they are unable to accurately match a desired target bitrate, which leads to packet loss and frame corruption when they exceed available capacity~\cite{karimi2023vidaptive}.

To overcome some of these limitations, several neural codecs have been proposed in recent years~\cite{swift, nas, srvc, Maxine, dvc}. These codecs use a low-quality video that is then enhanced using a Deep Neural Network (DNN). \namereparo\ differs from such prior neural codecs in two ways. First, these neural codecs still rely on temporal dependencies, wherein an undecodable frame can cause one or more subsequent frames to freeze. In contrast, \namereparo\ has no dependencies between encoded frames and hence the impact of a loss in one frame does not propagate to other frames. Second, none of the neural approaches use generative neural models that synthesize images from a few small pre-computed tokens. 

\noindent\textbf{Loss-Resilient Video Codecs.} 
Forward Error Correction (FEC) is a technique used in communication systems to recover lost data packets without retransmission. Instead of retransmission, redundant information sent by the sender is used by the receiver to reconstruct the original data. This is particularly important in real-time communication systems such as video conferencing, where retransmission of lost packets can cause unacceptable delays. Traditional FEC codes such as parity codes~\cite{begen2010rtp}, Reed-Solomon (RS) codes \cite{reed1960polynomial}, and fountain codes \cite{mackay2005fountain} are all block codes that are optimal for random losses, where packets are lost independently. Recently, researchers have proposed using streaming codes for FEC (e.g., Tambur~\cite{tambur}), achieving better loss recovery capabilities than block codes for bursty losses, where several packets over one or more consecutive frames are lost.

\red{GRACE \cite{cheng2023grace} proposed a neural video codec that can tolerate packet loss. However, there are two key differences with our approach. First, GRACE does not include an explicit loss recovery network. It improves the loss-resilience of an existing neural decoder (DVC~\cite{dvc}) by randomly masking information (i.e., using dropout) during its training. Unlike our loss recovery network, GRACE's decoder cannot leverage multiple received frames to reconstruct the missing information. Second, the DVC code underlying GRACE is based on Delta coding which creates temporal dependencies between encoded frames. Therefore, errors in one decoded frame due to packet loss propagate to subsequent frames even if they incur no loss. On the other hand, Reparo encodes each frame independently, and thus does not have error propagation.}


\NewPara{Generative Neural Networks.}
In recent years, there has been significant progress in the development of generative models, which can create text, audio, images, and videos indistinguishable from those created by humans \cite{devlin2018bert, liu2023audioldm, rombach2022high, li2022mage}. These models use knowledge of the target domain to generate content under certain conditions. For example, a text generative model can generate a paragraph conditioning on text prompts \cite{devlin2018bert}, and an image generative model can produce an image using only a partial view \cite{li2022mage}.

To enhance the use of domain knowledge, many recent visual generative models have adopted a two-stage design \cite{OordVK17, razavi2019generating, chang2022maskgit, yu2021vector, LeeKKCH22}. First, they learn to represent the target domain using a visual token codebook. Each visual token corresponds to a patch in the image and the codebook serves as a high-level abstraction of the visual world. Generation is then performed in this token space, similar to text generative models. These models have shown impressive performance in image generative tasks, such as text-to-image synthesis \cite{rombach2022high, chang2023muse} and image editing \cite{li2022mage}.

Given these capabilities, generative models are well-suited for loss-resilient video conferencing. Our work is the first to apply such advances to synthesize video conferencing frames when packet losses occur. By conditioning on (i.e., prompting with) the received data, our method can generate video frames identical to the original frames, achieving loss-resilient video conferencing.

\section{\namereparo Design}
\label{sec:design}

\subsection{Overview}
\label{sec:design overview}
\namereparo is a generative loss-resilient video codec specifically designed for video conferencing. As shown in \Fig{overview}, \namereparo~consists of five parts: (1) an \emph{encoder} that encodes the RGB video frame into a set of tokens, (2) a \emph{packetizer} that organizes the tokens into a sequence of packets,  (3) a \emph{bitrate controller} that adaptively drops some fraction of the packetized tokens to achieve a target bitrate, (4) a \emph{loss recovery module} that recovers the missing tokens in a frame based on the tokens received by the frame deadline, and (5) a \emph{decoder} that maps the tokens back into an RGB frame. We call the encoder-decoder combination in \namereparo~its neural codec, while the rest of the components help with loss recovery atop the codec. The encoder, packetizer, and bitrate controller are situated at the transmitter side, while the loss recovery module and decoder operate at the receiver side. We describe these modules in detail below.


\subsection{\namereparo Components}
\label{sec:components}

\subsubsection{The Neural Codec: Encoder and Decoder}\label{sec:codec}

In contrast to prior work on loss-resilient video conferencing, which utilizes traditional codecs with FEC-based wrappers, \namereparo\ employs its own codec based on the concept of a tokenizer. Tokenizers are commonly used in generative models to represent images using a learned codebook of tokens. Instead of generating images pixel by pixel, images are divided into patches, and each patch's features are mapped to a specific token in the codebook. This reduces the search space of generative models since the number of tokens in an image is much smaller than the number of pixels. Each token represents a vector in feature space. By training a neural network to identify a small number of feature vectors that can best generate all images in the training dataset, a set of tokens is selected for the codebook.

We observe that tokenizers naturally fit the requirements of a codec since they allow us to compress frames in a video by expressing them as a set of tokens, which can be transmitted as indices without the need to transmit the actual tokens. Since the transmitter and receiver share a codebook, the receiver can recover the original frames by looking up the token indices in its codebook and decoding them to the original frames. Further, since each frame is compressed independently of other frames based only on its own token indices, losses in one frame do not affect other frames.

\begin{figure}[t]
\begin{center}
\includegraphics[width=1.0\linewidth]{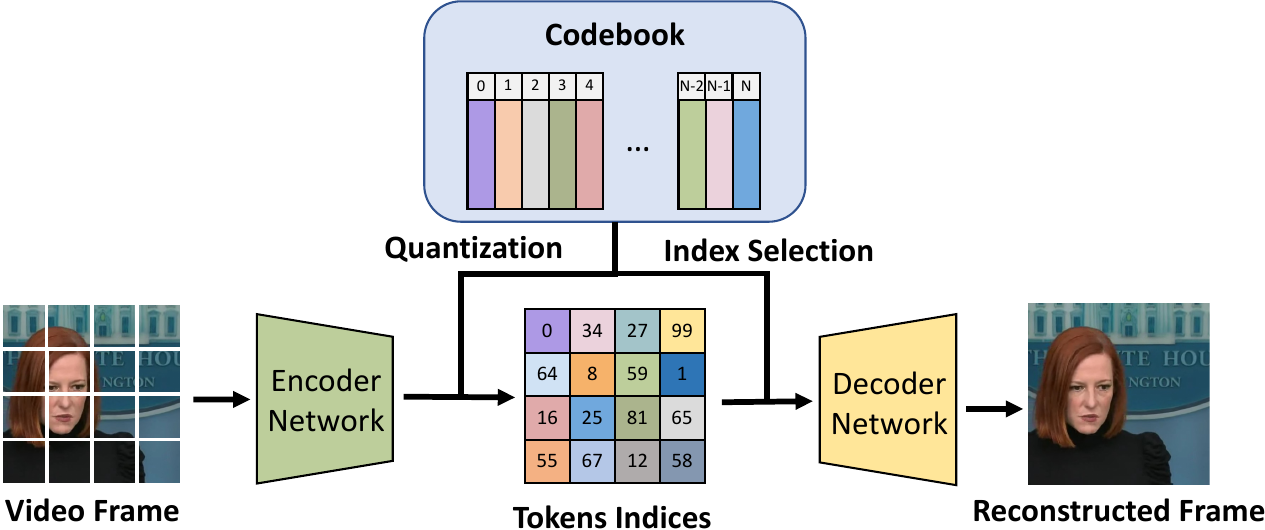}
\vspace{-5pt}
\end{center}
    \caption{Token-based neural codec. The encoder converts patches from video frames into features and uses a codebook to quantize the features into tokens by finding the nearest neighbor of each feature in the codebook. The decoder then uses the tokens to reconstruct the video frame.}
    \label{fig:vqgan}
\vspace{-5pt}
\end{figure}


We use a tokenizer called VQGAN~\cite{esser2021taming}, which consists of an encoder, a decoder, and a codebook (see \Fig{vqgan}). The encoder is a convolutional neural network (CNN) that takes patches in an image and maps each one of them to the nearest neighbor vector in the codebook, i.e., the nearest token. The decoder is also a CNN that takes a concatenation of tokens that represent an image and reproduces the original image. 

The compression achieved by VQGAN depends on two of its parameters: the number of tokens used for each frame, and the size of the codebook. Since the image is divided into patches, each mapped to a token, the number of tokens dictates the size of each patch within an image. As the number of tokens is increased, the smaller each patch becomes. More tokens allow a more fine-grained reconstruction as it is easier for a token to represent a smaller patch. However, since we transmit token indices from the sender to the receiver, more tokens means more bits for transmitting all of their indices, and reduces the compression factor. Similarly, a larger codebook enables a more diverse set of features to choose from for each token, but requires more bits to represent each token index. Thus, both of these parameters lead to different tradeoffs for the achieved bitrate and visual quality. We show this in \Fig{variants}.


\subsubsection{The Packetizer}
\label{sec:packetization}
\begin{figure}[t]
\begin{center}

\includegraphics[width=1.0\linewidth]{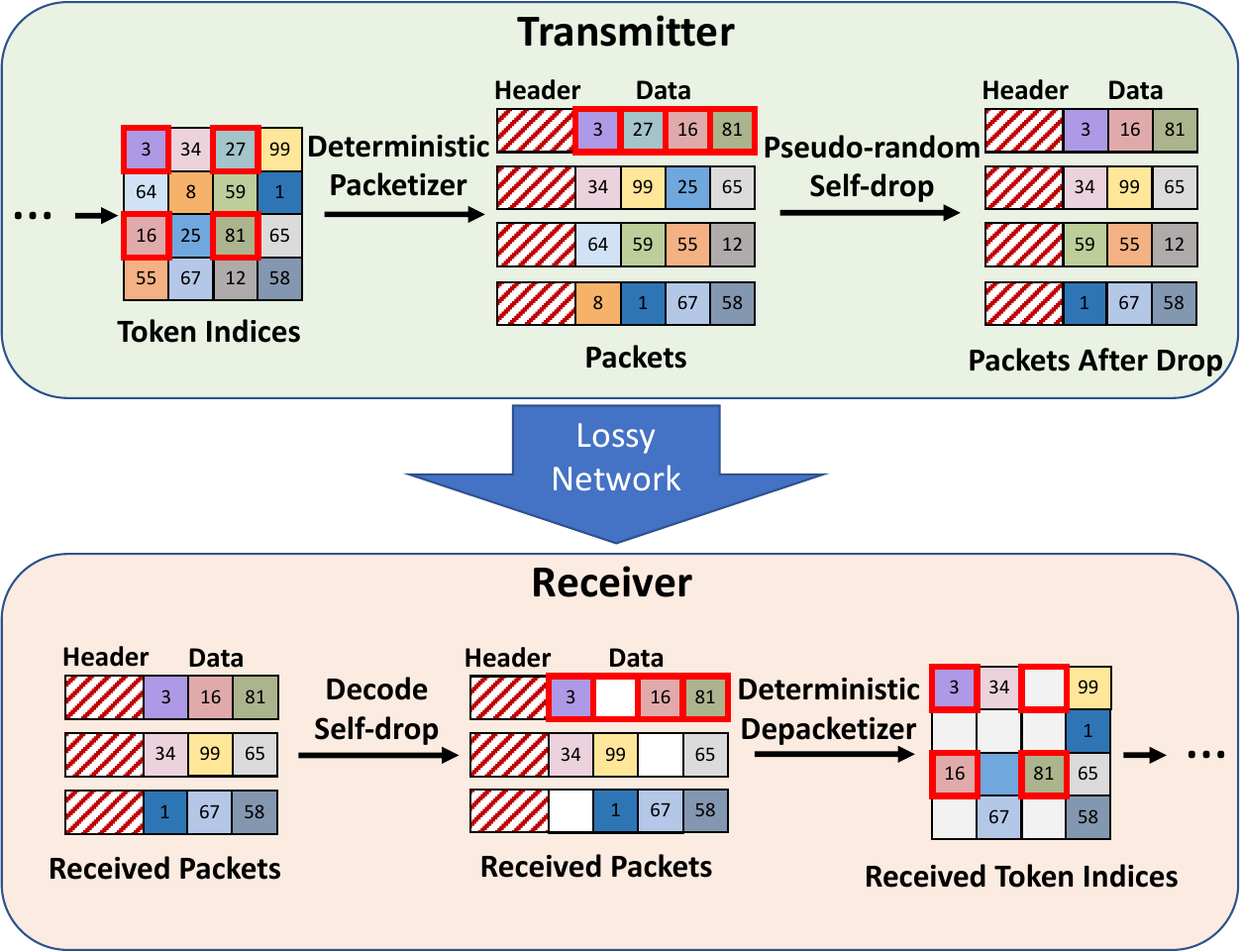}

\end{center}
\vspace{-5pt}
\caption{The transmitter first uses a deterministic packetizer to wrap image tokens into packets. Then a bitrate controller drops some tokens in each packet to adapt to the target bitrate. The receiver first decodes which tokens are dropped by the bitrate controller. It then depacketizes the received packets to extract the received token indices with the lost tokens identified.}\label{fig:packetization}
\vspace{-5pt}
\end{figure}

After encoding the original image into tokens, \namereparo divides them into several packets to prepare them for transmission. The packetization strategy is designed to avoid placing adjacent tokens in the same packet since the closest tokens in the image space are the most helpful for recovery when a token is lost.

In \Fig{packetization}, the first step in the green box labeled Transmitter shows our token wrapping strategy for an example with 4$\times$4 tokens that are split into 4 packets. The packet index of the token at position $(i, j)$ is $2\cdot(i \mod 2) + j \mod 2$. Tokens in each packet are ordered first by their row index, then by their column index in ascending order. This is just one of many ways to wrap tokens into packets while avoiding placing adjacent tokens in the same packet. The strategy needs to be deterministic so that the receiver can place the received tokens in the appropriate position in the frame before trying to recover the missing tokens.
Each packet has a header that includes its frame index, packet index, and packet size so that the receiver can identify which frame the tokens belong to and how many packets that particular frame has.


\subsubsection{The Bitrate Controller}
\label{sec:bitrate control}
Video conferencing applications often need to adjust their bitrate in response to network congestion. In prior work, this was achieved by altering the extent of compression to meet the desired bitrate. In contrast, \namereparo\ can easily adapt its bitrate by dropping tokens, as it is highly resilient to lost tokens and degrades gracefully with increasing loss rates. We call this ``self-dropping'' since \namereparo chooses to drop tokens on its own even before transmitting them. Remarkably, \namereparo\ can tolerate up to 50\% token loss with only a minimal impact on video PSNR, as demonstrated in \Fig{fig1}. 
In practice, \namereparo chooses the tokens it drops deterministically based on the frame index and packet index (\Fig{packetization} top row right). This is to ensure that the receiver can easily identify which token locations were dropped based simply on the frame and packet index in the received packet's header. With this information, the receiver can decode (\Fig{packetization} bottom row left) the locations of the tokens removed by the bitrate controller.

\subsubsection{Loss Recovery Module}
\label{sec:loss recovery}
\begin{figure}[t]
\begin{center}

\includegraphics[width=1.0\linewidth]{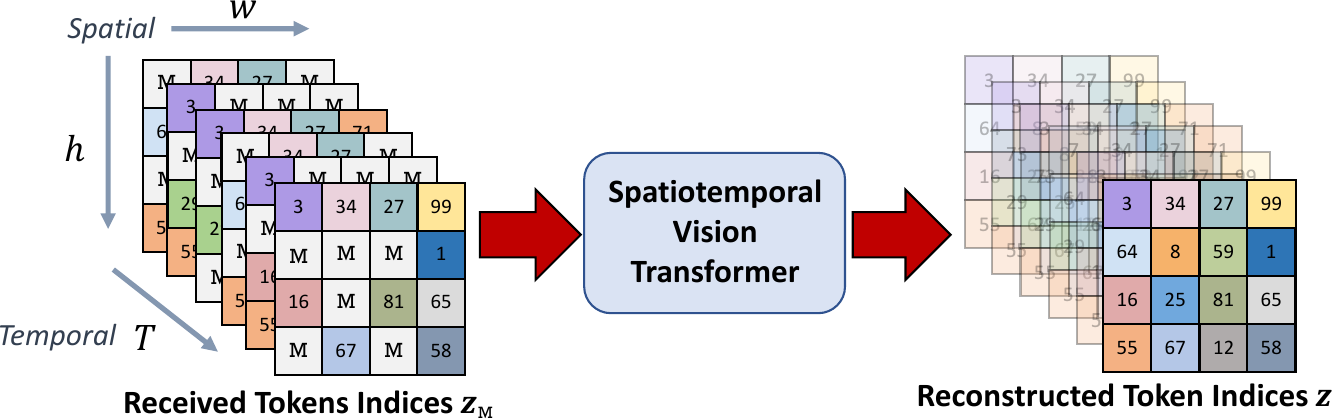}

\end{center}
\vspace{-5pt}
\caption{Loss recovery module. It uses a neural architecture based on a spatio-temporal vision transformer to generate any lost tokens using the learned knowledge of how people and objects look, along with the received tokens in the current and recent frames.}\label{fig:mage}
\vspace{-5pt}
\end{figure}

The key ingredient for \namereparo to carry out loss recovery is a deep generative model that leverages received tokens and video conferencing domain knowledge to generate lost tokens. For instance, the generative model can synthesize all tokens associated with a particular human face based on a subset of those tokens. Similarly, it can produce the token corresponding to a moving hand conditioned on the tokens from previous frames. In the following sections, we provide a comprehensive description of the architecture, training procedure, and inference algorithm of our loss recovery module.

\noindent\textbf{Network Architecture.} The loss recovery module is a neural network. It takes as input the received tokens organized according to their positions in the original frame. Lost tokens are expressed with a special token called the Mask token, \texttt{[M]}, as shown in \Fig{mage}. It also takes as input the tokens from the past $T$ frames, which provide the context for the scene. 

We use a common neural network architecture called Vision Transformer (ViT)~\cite{vit}. Transformers have gained widespread popularity in computer vision and natural language processing for predicting missing image patches or words \cite{devlin2018bert,vit,MAE}. The ViT employs an attention module in each layer to aggregate information from all tokens in an image. To predict a missing token, the attention module uses the received tokens and weighs them by their relevance to the missing token. The relevance is computed by performing a softmax over the dot product of each token with every other token. To extend the standard vision transformer structure to video clips, we use a spatio-temporal attention module \cite{arnab2021vivit}. In each transformer block, we perform attention over the time dimension $T$ (across adjacent frames) and then over the space dimension $h\times w$ within a frame. This enables our loss recovery module to exploit both spatial information from the same frame and temporal information across consecutive frames. Specifically, to generate a missing token, the module can use the nearby tokens in both space and across frames, as those tokens have a strong correlation with the missing token. Performing attention over time and space sequentially significantly reduces the computational cost: attention over both space and time simultaneously requires $O(T^2h^2w^2)$ of GPU memory, while attention first over time and then over space requires only $O(Th^2w^2 + T^2hw)$ of GPU memory.

Leveraging temporal information incurs some overhead as the last few frames need to be held in memory to decode the next frame. Hence, we limit the temporal dependency to a maximum of 6 frames. It is worth noting that using tokens from previous frames for loss recovery does not cause \namereparo to stall like traditional codecs due to undecodable frames. Specifically, the spatio-temporal ViT utilizes the six previous frames while decoding the current frame, allowing reuse of received tokens across frames to achieve a better bitrate and loss rate. Every frame is generated and decoded regardless of the previous frame's generation result and based solely on the actually received tokens of previous frames. If more tokens are lost in the previous frames, the quality of the current frame's generation may be poorer, but \namereparo will never stop generating or decoding, unlike classical codecs. We provide detailed information about our spatio-temporal ViT structure in \Sec{loss-recovery-implementation}.

It is worth highlighting the difference between our loss resilience and all past work. Traditionally, loss resilience is achieved by encoding frames together and adding FEC, at the transmitter. In contrast, our generative approach allows frames to be encoded \emph{independently} at the transmitter without FEC. The receiver however decodes each frame holistically, looking both at its received tokens and tokens from past frames to generate the missing tokens.

\noindent\textbf{Training the Network.} The goal of the training is to ensure the resulting neural network can recover from both network packet losses and tokens self-dropped by the bitrate controller to achieve a particular target bitrate.

Thus, during training, we simulate both types of losses and optimize the network weights to recover the original tokens. 
Specifically, we simulate the packetization process, and in each iteration, we randomly sample a self-drop ratio $r_d$ from 0 to 0.6. Based on $r_d$, a certain fraction of tokens are dropped from each packet. Then, a packet drop rate $r_p$ is randomly selected from 0 to 0.8, and packets (and all their tokens) are dropped based on the selected packet drop rate. At the receiver, the tokens that have been received are identified based on frame and packet indices. The missing tokens, whether dropped due to self-drops or packet loss, are replaced with a learnable mask token \texttt{[M]} (\Fig{mage}). This ensures that the input sequence length to the model is fixed regardless of the number of dropped tokens, which is a requirement for ViT. The resulting tokens combined with positional embeddings that provide spatial and temporal location information for each token (including the mask tokens) are then provided as the input of the ViT module. The output of the ViT module is a complete $h\times w \times T$ grid with generated or original tokens in their proper positions (where $T$ represents the number of past frames), but we only use the last frame's tokens to reconstruct the original frame using the codec decoder. Below we describe the loss function used in the training. 

\noindent\textbf{Reconstructive Training Loss.} Let $z=[z_{ijk}]_{i=1, j=1, k=1}^{h, w, T}$ denote the latent tokens from the encoder, and $M=[m_{ijk}]_{i=1, j=1, k=1}^{h, w, T}$ denotes a corresponding binary mask indicating which tokens are missing in the last $T$ frames. The objective of the training is to reconstruct the missing tokens from the available tokens. To accomplish this, we add a cross-entropy loss between the ground-truth one-hot tokens and the output of the loss recovery network. Specifically,

\vspace{-5pt}
\begin{equation}
\mathcal{L}_{reconstructive} = -\mathbb{E}{z} \big(\sum_{\forall i,j,k, m_{ijk}=1, k=T} \log p(z_{ijk}|z_{M})\big),
\end{equation}
where $z_M$ represents the subset of received tokens in $z$, and $p(z_{ijk}|z_{M})$ is the probability distribution over the codebook for position $(i, j)$ in the $k$-th frame predicted by the reconstruction network, conditioned on the input received tokens $z_M$. As is common practice, we only optimize this loss on the missing tokens of the last frame. Optimizing the loss on all tokens reduces reconstruction performance, as previously observed~\cite{MAE}. Detailed training schemes are included in \Sec{implementation}.

\noindent\textbf{Inference Routine} As the deadline for displaying each frame is hit every 33 ms for 30 fps, we aggregate all received packets for the current frame (as identified by the frame and packet indices) and regard all unreceived packets as lost. Once we place the received tokens in their respective positions corresponding to $h \times w$ patches in the frame, we can determine the exact locations of the missing tokens. We use all the tokens \emph{received} from the previous $T = 6$ frames to perform spatio-temporal loss recovery.

For each token position $(i, j)$ in the current frame, we use $p(z_{ij}|z^{M})$, the probability distribution of the predicted token given the received tokens, to choose the token with the highest probability as the reconstructed token. The resulting grid of reconstructed tokens is fed into the neural decoder to generate the RGB frame for display.






\section{Evaluation}
\label{sec:eval}
We evaluate \namereparo\ and compare to several baselines. 
We describe the baselines and experimental setup in \Sec{eval setup}. We evaluate baselines and \namereparo under network scenarios with random packet loss in \Sec{results}, and under packet losses induced by a rate-limited bottleneck link in \Sec{capacity-results}. We discuss \namereparo's parameter choices, latency overheads, and qualitative results in \Sec{eval others}.

\subsection{Experiment Setup}
\label{sec:eval setup}

\noindent\textbf{Baselines.}
\red{\textbf{ULPFEC} and \textbf{flexFEC} are two solutions included in WebRTC \cite{webrtc} to recover from audio and video packet loss. \textbf{Tambur}~\cite{tambur} is a recent streaming-codes based FEC solution atop the VP9 video codec \cite{vp9} that has been shown to perform better than classical block-based FEC techniques.  
Following the original paper, Tambur is implemented in the Ringmaster video conferencing platform ~\cite{ringmaster}. 
We set Tambur's latency deadline $\tau$ to 3 frames, and the bandwidth overhead of all baselines to be approximately 50\%. The frame rate is set to \SI{30}{fps} and the video resolution to 512$\times$512, which is typical for video conferencing.}

\noindent\textbf{Datasets.} For training the neural codec in \namereparo, we use a combination of three datasets: the FFHQ dataset (70,000 images) \cite{ffhq}, the CelebAHQ dataset (30,000 images) \cite{celeba}, and part of the TalkingHeads dataset ($\sim$25 hours of video) \cite{talkinghead}. These datasets comprise high-resolution human face images and videos, making them ideal for training \namereparo, 
which is aimed at improving video conferencing quality.
We exclusively use the TalkingHeads dataset for training the loss recovery module, as this module operates on video clips instead of images.

\red{For evaluating \namereparo and the baselines in the context of video conferencing applications, we combine the dataset from Gemino \cite{sivaraman2022gemino} and part of the TalkingHeads dataset \cite{talkinghead}, forming a large and diverse video conferencing validation dataset consisting of 5 hours of data from 412 video clips and 84 different subjects. We make sure that there is no overlap between the training and the validation set. Notably, our validation set is much larger and more diverse than prior works on video conferencing, such as Gemino \cite{sivaraman2022gemino} (25 videos from 5 people, 75 minutes in total), Tambur \cite{tambur} (20 videos, 200 minutes in total), and GRACE \cite{cheng2023grace} (60 videos, 745 seconds in total), demonstrating the generalization ability and effectiveness of \namereparo.}

\NewPara{Implementation.} 
We studied two FEC algorithms, ULPFEC and flexFEC, implemented in Google's implementation of WebRTC~\cite{webrtc} with VP8 codec.  The FEC rate is set to 50\% with packet retransmissions disabled, and the frame wait-time is set to \SI{150}{ms}. This wait-time parameter dictates the duration after which WebRTC stops to decode a frame that has been partially received. Our experimental setup includes a headless real-time video application, built on the WebRTC platform, which processes video input at a rate of \SI{30}{fps} at the sender's end and records the output to a file at the receiver's end. The network conditions between the sender and receiver were simulated using the Mahimahi network emulator~\cite{netravali2015mahimahi}. Within this environment, we constructed a loss shell akin to that used in Tambur, facilitating controlled network loss scenarios for our experiments. Tambur is evaluated using the Ringmaster platform with VP9 codec.

\red{Our neural codec and loss recovery modules were implemented in PyTorch, and they operate in real-time on two V100 GPUs, processing 30 fps 512x512 videos. One GPU was used for the transmitter and one GPU for the receiver. A V100 GPU is similar in performance to an Apple M2 Max GPU, which is integrated into a standard Macbook Pro laptop. To make a fair comparison, we use network traces from the WebRTC evaluation to evaluate the performance of \namereparo.}

\NewPara{Metrics.} We evaluate multiple metrics, \red{including peak signal-to-noise ratio (PSNR), structural similarity (SSIM)~\cite{ssim}, Learned Perceptual Image Patch Similarity (LPIPS)~\cite{lpips}, and percentage of non-rendered frames. PSNR, SSIM, and LPIPS are computed by comparing the displayed videos at the receiver to the original videos at the transmitter, using the PyTorch Image Quality (PIQ) library~\cite{kastryulin2022piq}.} Non-rendered frames are defined differently for baselines and \namereparo. For baselines, we compute the percentage of frames that are not played by the receiver due to packet loss on that frame itself or dependency on previously undecodable frames. For \namereparo, we define ``non-rendered frames'' as those frames with PSNR less than \SI{30}{dB}, as our scheme always tries to generate and render a frame. We have observed that for our dataset, VP8/9's PSNR rarely drops below \SI{30}{dB} unless there's an undecodable frame, so treating \namereparo's frames with PSNR below \SI{30}{dB} as ``non-rendered'' would favor baselines in comparisons. Non-rendered frames correlate well with standard quality-of-experience (QoE) metrics, as a large number of non-rendered frames can lead to video freezes and degrade QoE. All metrics are aggregated over all frames of the videos in our validation dataset. We compute bitrate by averaging the packet sizes (without TCP/IP headers) recorded in the network traces over the course of the entire video.

\noindent\textbf{Network Scenarios.} We consider two primary scenarios that can result in packet loss during transmission: (1) an unreliable network (\Sec{results}), and (2) a rate-limited link (\Sec{capacity-results}). In the first scenario, the network is unreliable and randomly drops packets due to poor conditions. To simulate this scenario, we use a GE loss channel. This channel transitions between a ``good'' state with a low packet loss rate and a ``bad'' state with a high packet loss rate, similar to Tambur's setup~\cite{tambur}.  
The probability of transitioning from the good state to the bad state and vice versa is 0.068 and 0.852, respectively. The probability of loss in the good state is 0.04. These parameters are set to mimic Tambur's evaluation, and are computed to approximate the actual statistics over a large corpus of traces from Microsoft Teams \cite{tambur}. We vary the probability of loss in the bad state to evaluate baselines and \namereparo's performance under different loss levels. Specifically, we set it to 0.25 to simulate a low loss level, 0.5 to simulate a medium loss level, and 0.75 to simulate a high loss level (the default value in Tambur's evaluation is 0.5). The statistics of the simulated bursty loss networks are included in \Tab{stats}. In the second scenario, we consider a fixed-rate link that drops packets once saturated. To simulate this, we use a FIFO queue with a fixed queue length of 6KB and a drain rate of 320 Kbps.

\begin{table}[h]
\vspace{-5pt}
\caption{Statistics of bursty lossy networks emulated using GE loss channel.}
\vspace{-15pt}
\label{tab:stats}
\begin{center}{
\resizebox{0.48\textwidth}{!}{
\begin{tabular}{l|c|c|ccccc}
\toprule
         & Low Loss Level & Medium Loss Level & High Loss Level \\
\midrule
Loss rate in bad state & 25\% & 50\% & 75\% \\
Time in bad state & 7.4\% & 7.4\% & 7.4\% \\
Avg. loss rate & 5.6\% & 7.4\% & 9.3\%  \\

\bottomrule
\end{tabular}
}}
\end{center}
\vspace{-10pt}
\end{table}


\NewPara{\namereparo Parameters.} We use a 512$\times$512 frame size and compress it into 32$\times$32 tokens. With a codebook size of 1024, each token requires 10 bits to represent its index. The codebook is trained once across the entire dataset and frozen during evaluation, eliminating the need to transmit it during video conferencing. A frame of tokens is split into 4 packets, with a packet header size of 4 bytes containing a 20-bit frame index, a 2-bit packet index, and a 10-bit packet size. Therefore, to send all the tokens of a frame, each packet requires 324 bytes, resulting in a default bitrate of 311.04 Kbps (at \SI{30}{fps}). The tokens in each packet can be dropped up to 50\% to match the target bitrate using the ``self-drop'' mechanism described in \Sec{bitrate control}. We can further control the bitrate (and visual quality trade-off) by using a different codebook and number of tokens per frame as shown in \Fig{variants}.

\begin{figure*}[t]
\begin{center}
\vspace{-20pt}
\includegraphics[width=0.95\linewidth]{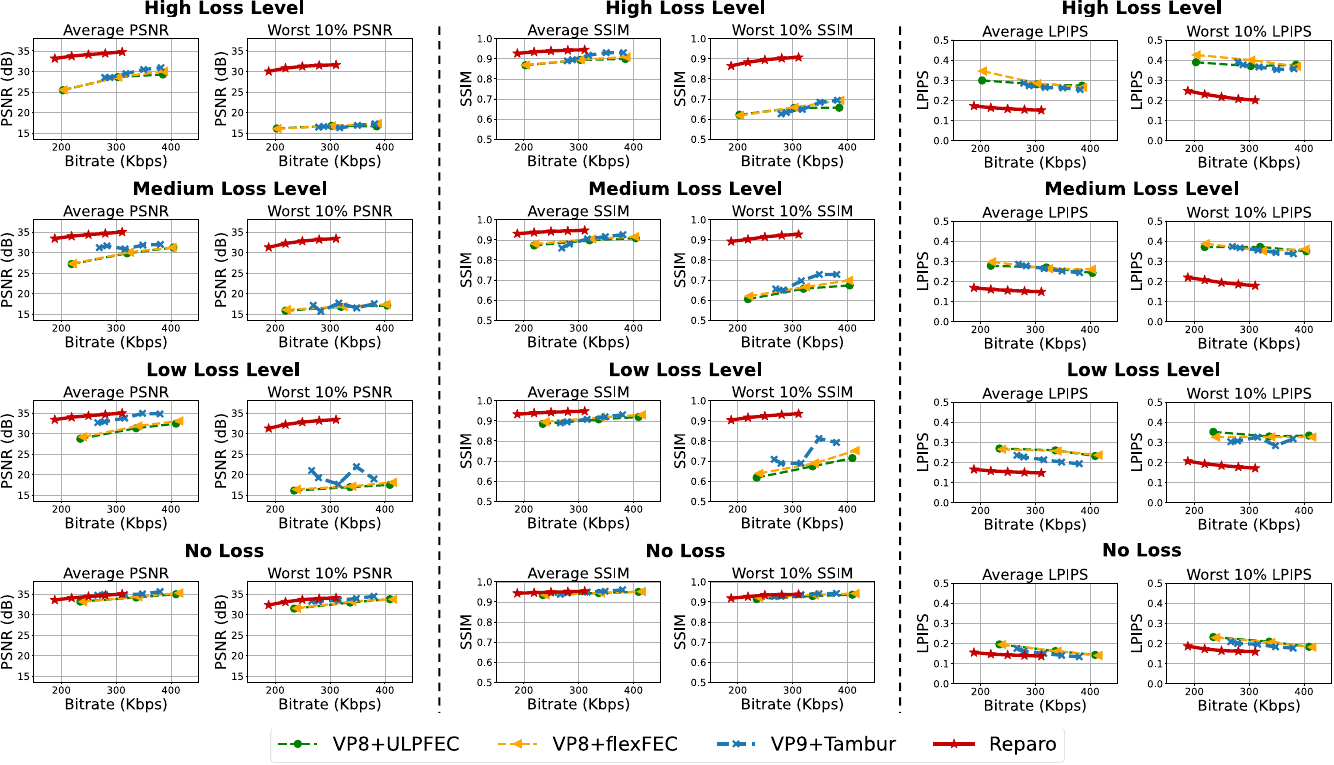}
\end{center}
\vspace{-10pt}
\caption{\red{We report the average and worst 10\% PSNR, SSIM and LPIPS of baselines and \namereparo~under different loss levels. PSNR and SSIM are the higher the better, and LPIPS is the lower the better. We vary the target bitrate of \namereparo~and baselines to cover different achieved bitrates. \namereparo's visual quality is significantly better than the baselines under all lossy conditions while achieving similar performance when there is no loss.}}\label{fig:fig1}
\vspace{-5pt}

\begin{center}
\includegraphics[width=0.9\linewidth]{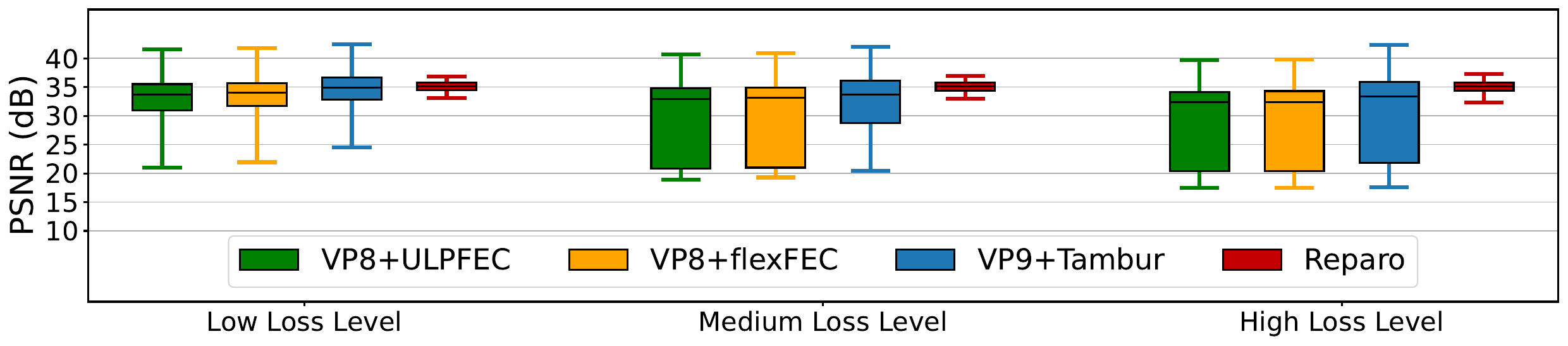}
\end{center}
\vspace{-15pt}
\caption{PSNR distribution across frames with baselines and \namereparo~under different packet loss rates (for a bitrate of $\sim$320 Kbps). The box denotes the 25\textsuperscript{th} and 75\textsuperscript{th} percentile PSNR, the line inside the box denotes the median PSNR while the whiskers denote average PSNR $\pm$ 1.5$\times$standard deviation. \namereparo maintains its PSNR within a narrow band around 35 dB regardless of the loss level while Tambur's worst frames drop to less than 20 dB PSNR at higher loss rates.}\label{fig:fig2}
\vspace{-10pt}
\end{figure*}

\subsection{Performance on Lossy Networks}
\label{sec:results}

\noindent\textbf{Visual Quality.} \red{We first compare the visual quality of video displayed using baselines and \namereparo~by evaluating the PSNR, SSIM, and LPIPS under different loss levels. We also vary the target bitrate to evaluate the performance of our method and baseline under different bitrate constraints. We measure the average performance over all frames and the worst 10\% frames, representing the overall video conferencing quality and the quality under poor network conditions. All metrics are aggregated across all frames in our evaluation corpus.}

As shown in \Fig{fig1}, \namereparo achieves better performance with smaller bitrates for all metrics and under all conditions. The poor performance of baselines is caused by freezes of displayed video: during a freeze, the video is stuck at the last rendered frame. In contrast, \namereparo maintains a high and stable performance even under high loss levels, thanks to two key design elements. First, \namereparo does not have any temporal dependency at the neural codec level. Encoding into and decoding from tokens occur on a frame-by-frame basis without any dependency on a previous frame. Thus, even if a frame's tokens are mostly lost, it could have a lower performance but will not affect subsequent frames whose tokens are received. Second, the loss recovery module uses a deep generative network that leverages domain knowledge of human faces to generate lost tokens. It will only fail to generate accurately if a very large portion of tokens is lost across packets over multiple frames, which is highly unlikely.

\label{sec:deep dive}
\begin{figure*}[t]
\begin{center}

\includegraphics[width=1.0\linewidth]{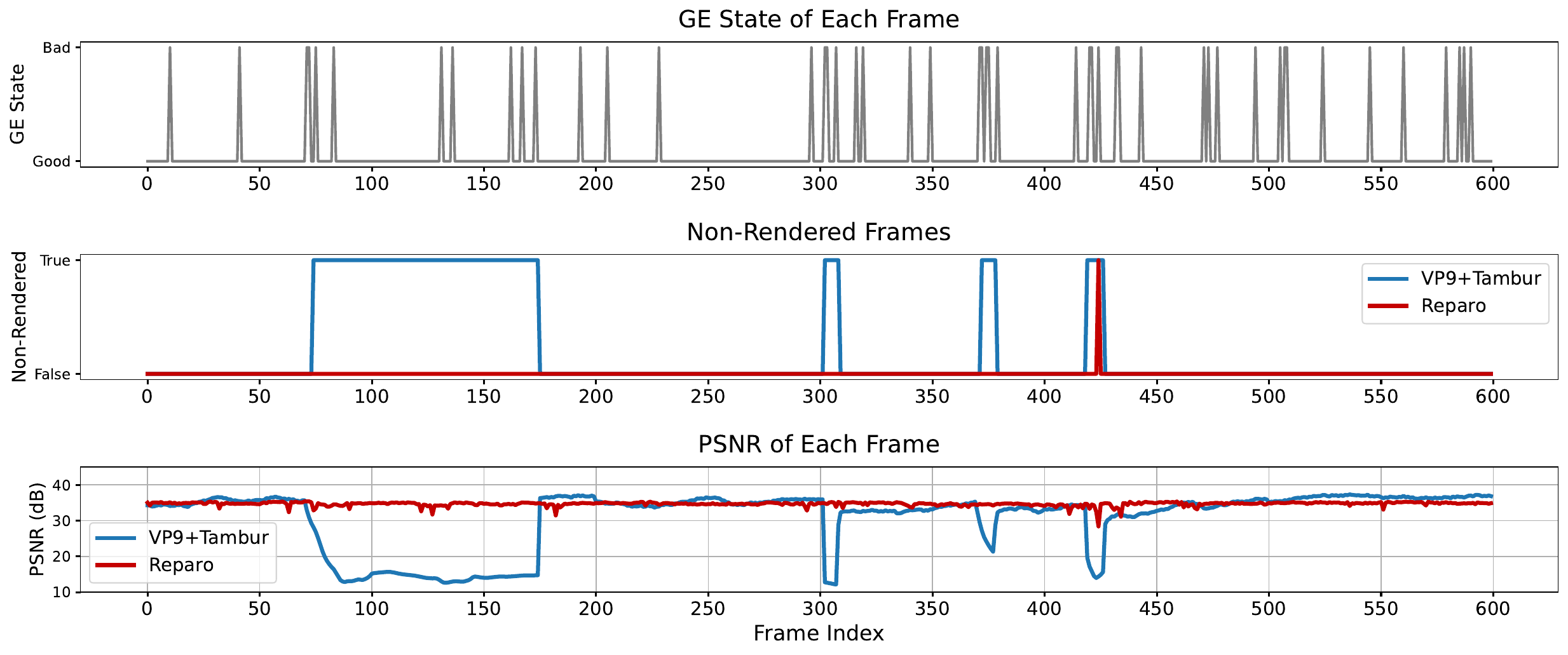}

\end{center}
\vspace{-15pt}
\caption{Timeseries comparing Tambur and \namereparo on one video and loss pattern. Tambur experiences short freezes every time a set of frames is lost with a corresponding decrease in PSNR. \namereparo continues rendering frames and its visual quality is a lot more stable throughout the interval.}\label{fig:fig3-1}
\vspace{-10pt}
\end{figure*}

\begin{figure}[t]
\begin{center}
\includegraphics[width=0.9\linewidth]{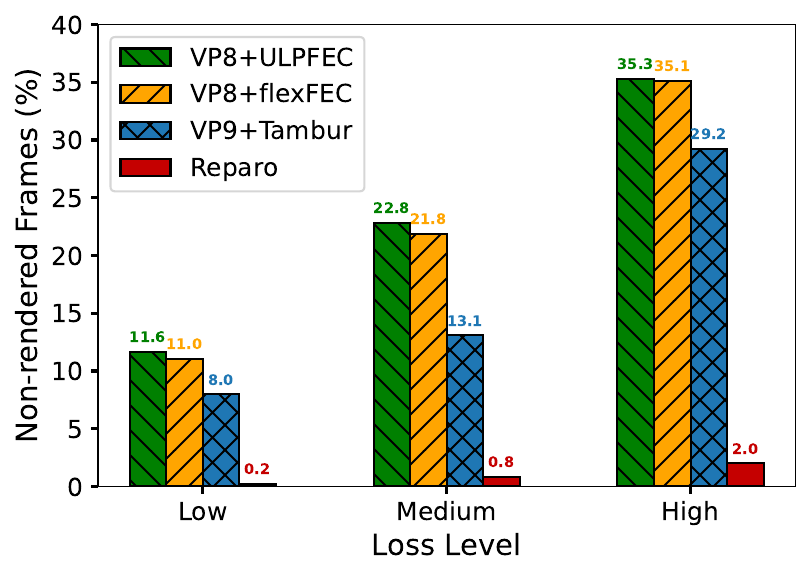}
\end{center}
\vspace{-15pt}
\caption{Comparison of percentage of non-rendered frames between \namereparo~and baselines. All baselines experience many more non-rendered frames than \namereparo at all loss levels.}\label{fig:nonrender}
\vspace{-15pt}
\end{figure}

We further show the distribution of frame PSNR values across the frames in our evaluation corpus with \namereparo~and baselines at $\sim$\SI{320}{Kbps} under different packet loss rates in the ``bad'' state of the GE channel. As shown in \Fig{fig2}, \namereparo's distribution and averages of PSNR values are more or less unaffected by the loss level. With \namereparo, almost 99\% of frames have PSNR values larger than \SI{30}{dB}. The variance of PSNR values across displayed frames is also much lower with \namereparo than the baselines, showing the stability of the quality of the displayed video across loss levels. Specifically, \namereparo's frame PSNR values are mostly between \SI{32.5}{dB} and \SI{37}{dB} (>90\%)). In contrast, as the loss level becomes higher, the baselines are more likely to experience video freezes, resulting in more frames with low PSNR. These results demonstrate that \namereparo is more robust and efficient at recovering from packet losses than current FEC schemes for video conferencing.

\noindent\textbf{Non-Rendered Frames.} Another commonly used metric for evaluating FEC approaches is the frequency of non-rendered frames, which can cause freezes in the displayed video. One advantage of \namereparo is that it never truly freezes: it always attempts to generate lost tokens and the frame, regardless of the packet loss rate. However, in extreme cases, it may still produce poor generated output. To provide a fair comparison, we define frames with a PSNR of less than \SI{30}{dB} as ``non-rendered frames'' for \namereparo, since we observed that VP8/9's PSNR rarely drops below \SI{30}{dB} unless frames are lost and the video stalls. We note that such a definition favors baselines in their comparisons with \namereparo since we do not penalize them for low-quality rendered frames.

We evaluated \namereparo and baselines at a similar bitrate ($\sim$\SI{320}{Kbps}) under various loss levels. As shown in \Fig{nonrender}, \namereparo nearly eliminates non-rendered frames under all loss levels, whereas all baselines have a noticeable number of non-rendered frames. This result further demonstrates \namereparo's effectiveness in displaying consistently high-quality videos even under severe packet losses, in contrast to current codecs and FEC schemes that cause extended freezes.

\begin{figure*}[t]
\begin{center}

\includegraphics[width=0.97\linewidth]{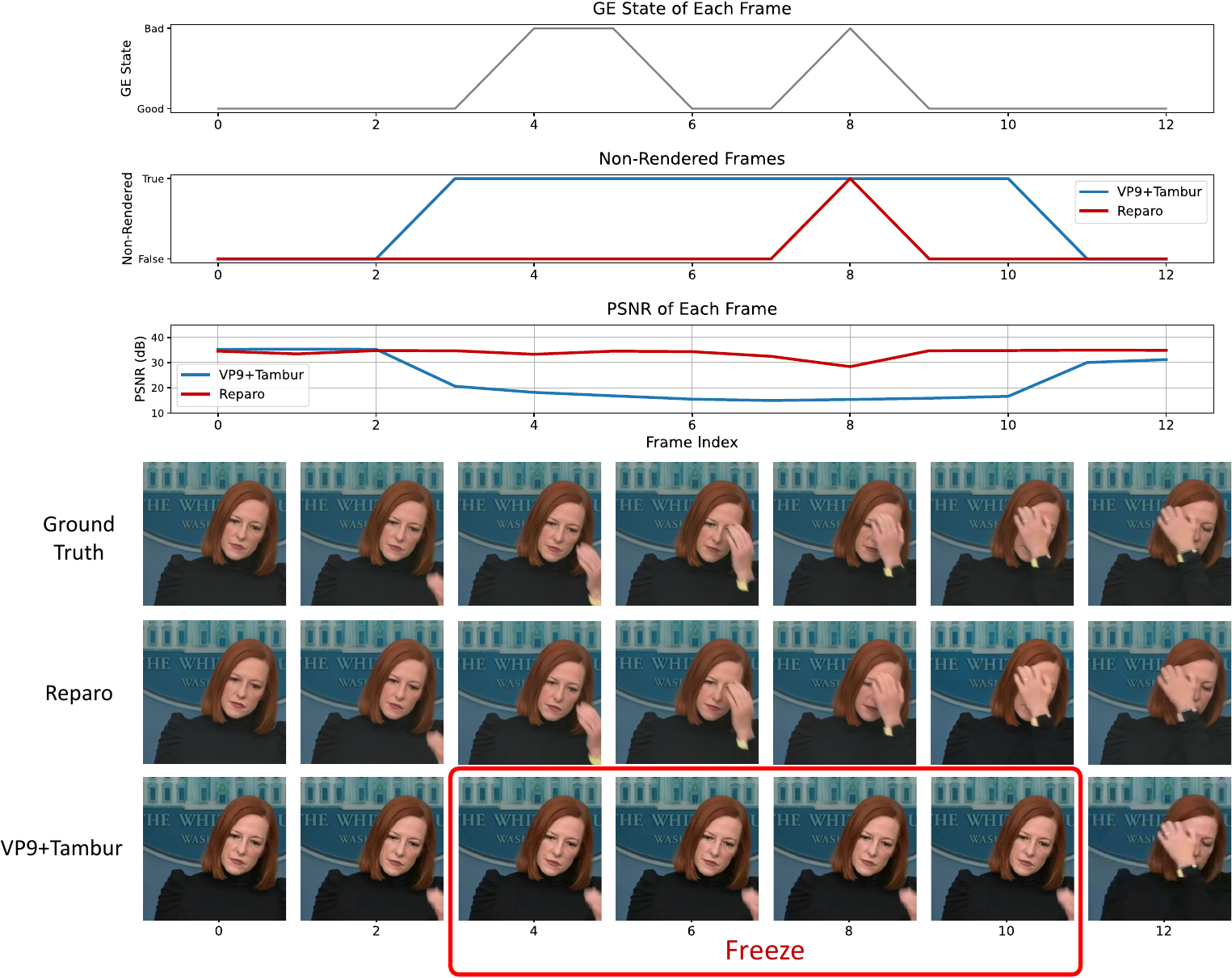}
\vspace{-5pt}
\end{center}
\caption{Qualitative results of Tambur and \namereparo~during a Tambur's short freeze of 8 frames. The GE loss channel is in a ``bad'' state at frames 4, 5, 6, and 8, causing packet losses for both VP9+Tambur and \namereparo. Tambur completely freezes from frames 3 to 10 because of lost packets, leading to very low PSNR. On the other hand, though \namereparo~experiences the same GE loss state as Tambur, it generates most of the frames and maintains a high PSNR. Even for the frame under 30 PSNR, it still produces reasonable output and tracks the hand movement accurately.}\label{fig:fig3-2}
\vspace{-15pt}
\end{figure*}

\noindent {\textbf{A Detailed Example.} 
To better understand \namereparo's benefits come from, we present a time series of loss patterns, non-rendered frames, and PSNR values for \namereparo and Tambur over a 30-second window in \Fig{fig3-1} for a particular video sequence. The loss level is set to medium (packet loss probability of 0.5 in the bad state). The sequence of lost frames starting at frame index 71 causes Tambur to experience an extended freeze between frame 72 and frame 177, even though many frames in that timeframe were not lost. This is due to temporal dependencies between video frames, where frames are compressed based on the differences between them. As a result, a lost frame can lead to subsequent undecodable frames (even when they're received successfully) until the encoder and decoder are reset using a keyframe. As expected, Tambur exhibits much lower PSNR ($\sim$\SI{15}{dB}) during that timeframe between frames 72 and 177. Tambur then forces the encoder to transmit a keyframe to resume the video stream. Subsequent frames' PSNR values go back to what they were prior to the freeze period. If such a keyframe is also lost (which is more likely because a keyframe is much larger than normal frames and contains more packets since it is compressed independently of its adjacent frames), it could cause long freezes that span several seconds.

In contrast, \namereparo is much more stable in PSNR and rarely experiences non-rendered frames, even during periods of loss. \namereparo may generate one or more frames with low PSNR if it loses many tokens, as happens at frame 424. However, its per-frame decoding structure ensures that its visual quality quickly recovers as tokens for future frames start coming in.

To gain a more comprehensive understanding of the effects of packet loss events, we examine a short freeze event of Tambur spanning 8 frames in greater detail and compare it to \namereparo in \Fig{fig3-2}. This figure shows lost frames, non-rendered frames, frame PSNR values as well as visuals of the displayed frames in that time interval. As depicted in the figure, part of the 3\textsuperscript{rd}, 4\textsuperscript{th}, and 5\textsuperscript{th} frames are initially lost, followed by the loss of the 8\textsuperscript{th} frame. Tambur does not render any frames between the 3\textsuperscript{rd} and 10\textsuperscript{th} frames, as is evident from the ``non-rendered'' frames line and the unchanged video frames in the visual strip beneath. Additionally, the forced keyframe (frame 11) and subsequent frame 12 have slightly lower PSNR due to the larger size of the keyframe, which typically has a lower quality to meet the target bitrate when compressed without any temporal dependency. In contrast, \namereparo does not experience such a prolonged freeze, as evidenced by the ``non-rendered frames'' row and the visual strip. Although \namereparo produces a lower PSNR frame at the 8\textsuperscript{th} frame, it rapidly recovers once later frames receive sufficient packets and tokens for high-quality generation.

\begin{figure}[t]
\begin{center}
\includegraphics[width=1.0\linewidth]{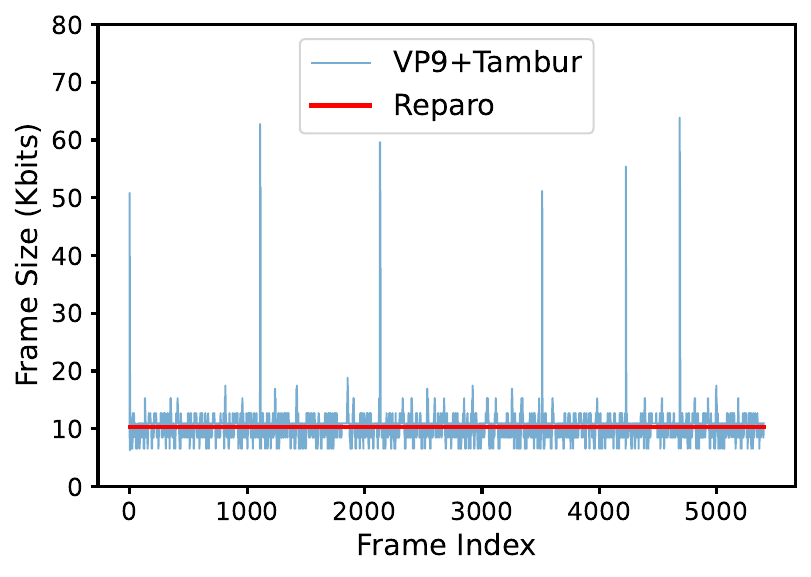}
\end{center}
\vspace{-5pt}
\caption{\small Per frame sizes of VP9+Tambur and \namereparo~for a 3 minute video. \namereparo maintains the same frame size across all frames while VP9 shows variance both across adjacent frames and across periodic keyframes that are large.}\label{fig:bitrate}
\end{figure}

\subsection{Performance on Rate-Limited Networks}
\label{sec:capacity-results}

In this section, we consider the packet loss caused by a \emph{rate-limited bottleneck link} when it saturates. One advantage of \namereparo is its ability to match and transmit at different target bitrates easily by simply varying the self-drop rate. This is because, unlike traditional temporal-dependent codecs, \namereparo does not need to transmit keyframes periodically. Instead, every frame is encoded into a set of tokens with the same size across frames. As shown in \Fig{bitrate}, VP9+Tambur needs to transmit a keyframe periodically, causing spikes in its per-frame sizes. Even the P-frames in VP9 show quite a bit of variance in their sizes. In contrast, \namereparo can always maintain a constant size across frames and consequently, constant bitrate because its neural codec encodes each frame with the same number of tokens.

Such a stable bitrate improves \namereparo's performance over fixed-capacity bottleneck links. To simulate such a link, we use a FIFO queue with a constant (drain) rate of $r$ Kbps. The size of the queue is set to $0.15\times r$, as such a queue will introduce a \SI{150}{ms} delay, which is the upper bound of industry recommendations for interactive video conferencing \cite{itu}.  Packets are queued first and drained at the desired link rate. When the FIFO queue becomes full, subsequent packets will be dropped. In \Fig{lossy-bucket}, we set $r$ to \SI{320}{Kbps} and show the average PSNR of \namereparo~and VP9+Tambur with different target bitrates for each codec. Note that the target bitrate for VP9+Tambur typically does not match the actual bitrate: it is the input parameter for the VP9 codec to encode a video. As a result, the actual bitrate of VP9+Tambur can be much larger than the target bitrate of VP9 depending on the encoding speed and quality parameters. Also, Tambur's parity packets typically introduce 50\% to 60\% bandwidth overhead, further inflating the actual bitrate of VP9+Tambur. For example, the \SI{75}{Kbps} target bitrate corresponds to an actual average bitrate of \SI{211}{Kbps}. As a result, we only vary the target bitrate supplied to VP9+Tambur up to \SI{200}{Kbps} because beyond that its actual bitrate with FEC overheads overshoots the link rate and causes a lot of packet drops. On the other hand, \namereparo's actual bitrate can exactly match the target bitrate.

\begin{figure}[t]
\begin{center}
\includegraphics[width=1.0\linewidth]{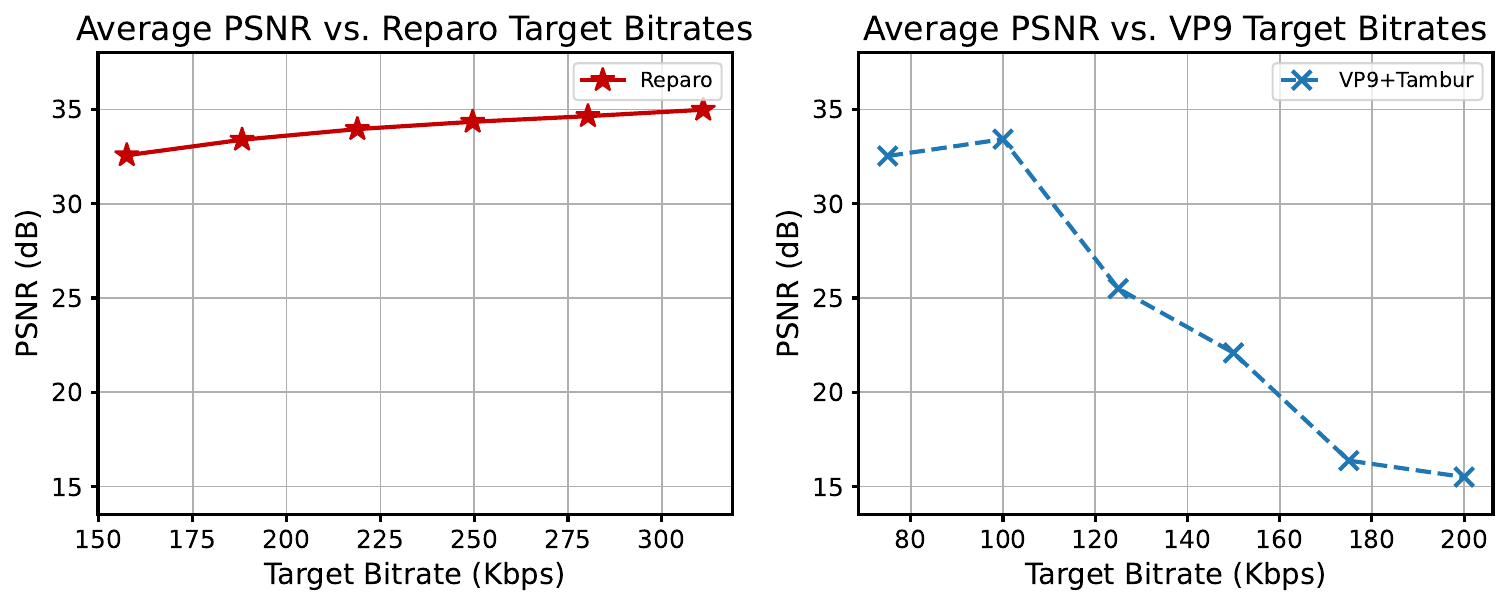}
\end{center}
\vspace{-5pt}
\caption{Average PSNR of \namereparo~and Tambur with different target bitrates for a fixed link capacity of 320 Kbps. \namereparo's average PSNR improves as the target bitrate is increased. However, VP9+Tambur starts experiencing loss in its fixed-size queue beyond a target bitrate of \SI{120}{Kbps} due to large keyframes that do not fit in the queue.}\label{fig:lossy-bucket}
\end{figure}

As shown in \Fig{lossy-bucket}, the average PSNR achieved by \namereparo increases as the target bitrate is increased. This is expected because fewer tokens are ``self-dropped'', allowing for better reconstruction. However, while the PSNR of Tambur initially increases as the target bitrate is increased, it begins to decrease when the target bitrate is set to \SI{120}{Kbps}. This occurs because even with a small target bitrate, the size of a keyframe across all its packets with VP9 can be much larger than the total number of bytes that the queue can hold. Consequently, many packets of this keyframe may be lost. Additionally, when a keyframe is lost, Tambur will force another keyframe, causing the queue to remain full and preventing any frames from being transmitted, resulting in a frozen video over long durations. As the target bitrate is increased further and this issue with keyframes becomes more pronounced, VP9+Tambur's average PSNR worsens.

In practice, congestion control protocols like GCC~\cite{gcc} are used to adapt the encoder's target bitrate based on network observations such as latency and loss. However, this experiment shows that choosing the appropriate target bitrate for Tambur is much more challenging than for \namereparo. For VP9+Tambur, the adaptation protocol must be conservative and operate in a lower bitrate range to limit packet drops. In contrast, \namereparo can continue to benefit from larger target bitrates as long as they are smaller than the link capacity, and the best performance is achieved by setting the target bitrate near the link capacity.

\begin{figure}[h]
\begin{center}

\includegraphics[width=0.95\linewidth]{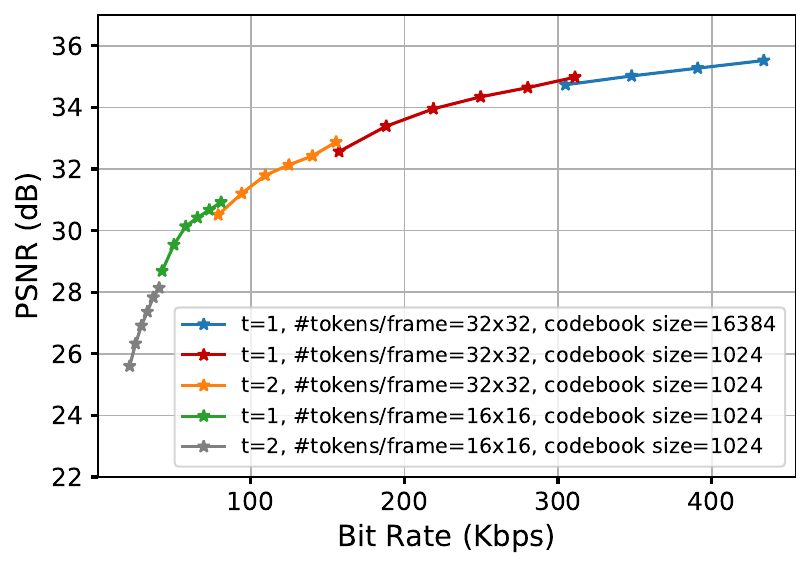}

\end{center}
\vspace{-15pt}
\caption{Variants of \namereparo~that operate in different bitrate regimes. \namereparo~ achieves different bitrates by varying the number of tokens per frame, its codebook size, and the number of frames jointly encoded.}\label{fig:variants}
\vspace{-15pt}
\end{figure}

\subsection{Other Results}
\label{sec:eval others}

\noindent\textbf{\namereparo Ablation Study.} To allow \namereparo to operate in different bitrate regimes, we can adjust its hyper-parameters. For example, we can compress $t$ adjacent frames into the same fixed size $h\times w$ tokens, which reduces the effective bitrate by a factor of $t$ at the cost of an additional latency of $t-1$ frames. We can also modify the number of residual blocks used in the encoder and decoder, which changes the number of tokens to represent a frame. More tokens per frame correspond to better PSNR and higher bitrate due to better representational power. We can also use different codebook sizes; larger codebook sizes produce higher PSNR at the cost of a larger bitrate. In \Fig{variants}, we show the PSNR-bitrate curve of \namereparo under different hyper-parameters with a low loss level, demonstrating that \namereparo~can be adapted to a large range of bitrates by varying the codec and loss recovery module trained under different hyper-parameters. For example, \namereparo can choose to encode two frames together at the cost of \SI{33}{ms} higher latency and achieve almost half bitrate (red curve and orange curve). \namereparo can also use a larger codebook to achieve higher PSNR at the cost of more bits needed to encode each token index (red curve and blue curve). By default, we use the middle red curve (t=1, number of tokens per frame=32$\times$32, codebook size=1024) for \SI{30}{fps} 512$\times$512 videos in our main experiments.

\begin{table}[h]
\vspace{0pt}
\caption{Latency breakdown for different parts of \namereparo. The encoder and packetization are at the transmitter side, while the loss recovery and decoder are at the receiver side.}
\vspace{-15pt}
\label{tab:latency}
\begin{center}{
\resizebox{0.48\textwidth}{!}{
\begin{tabular}{l|c|c|c|cccc}
\toprule
        & Encoder & Packetization & Loss Recovery & Decoder \\
\midrule
Latency (ms) & 14.1$\pm$0.1 & 0.5$\pm$0.009 & 17.8$\pm$1.0 & 13.1$\pm$0.3 \\

\bottomrule
\end{tabular}
}}
\end{center}
\vspace{-15pt}
\end{table}

\noindent\textbf{Latency.} In \Tab{latency}, we present the latency of different modules in \namereparo. The neural codec and loss recovery modules of \namereparo have higher encoding and decoding latencies compared to traditional FEC schemes, since they require heavy computation. For our implementation, the total inference delay incurred by \namereparo is \SI{45.5}{ms}. With typical network queuing delays of \SI{50}{ms}, the end-to-end delay of \namereparo is less than \SI{100}{ms}, which meets the industry recommendation of \SI{150}{ms} for maximum tolerable latency for interactive video applications \cite{itu}.

\noindent\textbf{Qualitative Results.} We also compare \namereparo~with Tambur qualitatively. Please refer to this \href{https://www.dropbox.com/scl/fi/xzocvdbfyqiv620pvo36y/qualitative.mp4?rlkey=hynicc9kxbc2bs10i9t5svpor&st=mwn8kzp3&dl=0}{\textcolor{red}{\texttt{link}}} for qualitative results.
\section{Limitations}
Although \namereparo\ offers several key advantages over past work, it also has some limitations. First, the current implementation of \namereparo\ is in PyTorch, and uses transformers which are computationally more intensive than traditional video codecs and FEC-based methods~\cite{vit,arnab2021vivit}. It requires GPUs equivalent to an Apple M2 Max GPU to operate in real time. This limits the range of devices on which \namereparo\ can be deployed, and the current implementation is not suitable for low-end devices such as smartphones or tablets. 
However, machine learning models can be sped up for edge devices using more efficient model architectures~\cite{mobilenet,efficientnet,architecture_search, rao2021dynamicvit}, hardware design~\cite{gale2020sparse, han2017ese, zhang2020sparch}, and techniques such as knowledge distillation~\cite{kd}. 
We leave an investigation of such optimizations to future work. We also note that over time more powerful GPUs are integrated in edge devices naturally paving the way for running complex neural networks on them. 
Second, \namereparo\ learns a dictionary of tokens specific to the domain of interest, which is video conferencing for the purpose of this paper. While \namereparo\ can potentially be extended beyond video conferencing to other domains, this will require learning a dictionary for each new domain. As such it is more specialized and less general than traditional FEC codecs. 


Despite these limitations, \namereparo\ represents a promising approach to loss-resilient video conferencing. Future research may focus on addressing these limitations and making \namereparo\ more accessible to a wider range of devices and different video-based applications.

\vspace{-2pt}
\section{Conclusion}

We present \namereparo, a novel loss-resilient generative video conferencing architecture that uses generative deep learning models to reconstruct missing information without sending redundant packets or relying on retransmissions. Instead, the receiver reconstructs missing information using its knowledge of how visual objects look and relate to each other. Our approach offers several advantages, including maintaining a constant bit rate, easy adaptation to any target bitrate, and one-way communication between the transmitter and receiver. We evaluate \namereparo\ on a large and diverse corpus of publicly available video conferencing videos and show that it consistently outperforms multiple FEC baselines including Tambur, a state-of-the-art loss-resilient video conferencing platform based on streaming FEC. \namereparo\ significantly improves over them under different loss levels, while also mostly eliminating video freezes. Our approach presents a promising solution to the challenges of real-time video conferencing applications, and we believe it opens up exciting possibilities for further research in this area.

\label{beforerefs} 

\clearpage
\bibliographystyle{plain}
\bibliography{main.bib}

\begin{thebibliography}{10}

\bibitem{av1}
{AV1} bitstream \& decoding process specification.
\newblock \url{http://aomedia.org/av1/specification/}.

\bibitem{ringmaster}
{Ringmaster}.
\newblock \url{https://github.com/microsoft/ringmaster}.

\bibitem{webrtc}
{WebRTC}.
\newblock \url{https://webrtc.googlesource.com/src}.

\bibitem{arnab2021vivit}
Anurag Arnab, Mostafa Dehghani, Georg Heigold, Chen Sun, Mario Lu{\v{c}}i{\'c},
  and Cordelia Schmid.
\newblock Vivit: A video vision transformer.
\newblock In {\em Proceedings of the IEEE/CVF international conference on
  computer vision}, pages 6836--6846, 2021.

\bibitem{vp8}
Jim Bankoski, Paul Wilkins, and Yaowu Xu.
\newblock Technical overview of {VP8}, an open source video codec for the web.
\newblock In {\em 2011 IEEE International Conference on Multimedia and Expo},
  pages 1--6. IEEE, 2011.

\bibitem{begen2010rtp}
Ali Begen.
\newblock Rtp payload format for 1-d interleaved parity forward error
  correction (fec).
\newblock Technical report, 2010.

\bibitem{boyce1999packet}
Jill~M Boyce.
\newblock Packet loss resilient transmission of mpeg video over the internet.
\newblock {\em Signal Processing: Image Communication}, 15(1-2):7--24, 1999.

\bibitem{gcc}
Gaetano Carlucci, Luca De~Cicco, Stefan Holmer, and Saverio Mascolo.
\newblock Analysis and design of the google congestion control for web
  real-time communication (webrtc).
\newblock In {\em Proceedings of the 7th International Conference on Multimedia
  Systems}, pages 1--12, 2016.

\bibitem{chang2023muse}
Huiwen Chang, Han Zhang, Jarred Barber, AJ~Maschinot, Jose Lezama, Lu~Jiang,
  Ming-Hsuan Yang, Kevin Murphy, William~T Freeman, Michael Rubinstein, et~al.
\newblock Muse: Text-to-image generation via masked generative transformers.
\newblock {\em arXiv preprint arXiv:2301.00704}, 2023.

\bibitem{chang2022maskgit}
Huiwen Chang, Han Zhang, Lu~Jiang, Ce~Liu, and William~T Freeman.
\newblock Maskgit: Masked generative image transformer.
\newblock In {\em Proceedings of the IEEE/CVF Conference on Computer Vision and
  Pattern Recognition}, pages 11315--11325, 2022.

\bibitem{cheng2023grace}
Yihua Cheng, Ziyi Zhang, Hanchen Li, Anton Arapin, Yue Zhang, Qizheng Zhang,
  Yuhan Liu, Xu~Zhang, Francis~Y. Yan, Amrita Mazumdar, Nick Feamster, and
  Junchen Jiang.
\newblock Grace: Loss-resilient real-time video through neural codecs, 2023.

\bibitem{swift}
Mallesham Dasari, Kumara Kahatapitiya, Samir~R. Das, Aruna Balasubramanian, and
  Dimitris Samaras.
\newblock Swift: Adaptive video streaming with layered neural codecs.
\newblock In {\em 19th USENIX Symposium on Networked Systems Design and
  Implementation (NSDI 22)}, pages 103--118, Renton, WA, April 2022. USENIX
  Association.

\bibitem{devlin2018bert}
Jacob Devlin, Ming-Wei Chang, Kenton Lee, and Kristina Toutanova.
\newblock Bert: Pre-training of deep bidirectional transformers for language
  understanding.
\newblock {\em arXiv preprint arXiv:1810.04805}, 2018.

\bibitem{vit}
Alexey Dosovitskiy, Lucas Beyer, Alexander Kolesnikov, Dirk Weissenborn,
  Xiaohua Zhai, Thomas Unterthiner, Mostafa Dehghani, Matthias Minderer, Georg
  Heigold, Sylvain Gelly, et~al.
\newblock An image is worth 16x16 words: Transformers for image recognition at
  scale.
\newblock In {\em Int. Conf. on Learning Representations (ICLR)}, 2021.

\bibitem{architecture_search}
Thomas Elsken, Jan~Hendrik Metzen, and Frank Hutter.
\newblock Neural architecture search: A survey.
\newblock {\em The Journal of Machine Learning Research}, 20(1):1997--2017,
  2019.

\bibitem{esser2021taming}
Patrick Esser, Robin Rombach, and Bjorn Ommer.
\newblock Taming transformers for high-resolution image synthesis.
\newblock In {\em Proceedings of the IEEE/CVF conference on computer vision and
  pattern recognition}, pages 12873--12883, 2021.

\bibitem{gale2020sparse}
Trevor Gale, Matei Zaharia, Cliff Young, and Erich Elsen.
\newblock Sparse gpu kernels for deep learning.
\newblock In {\em SC20: International Conference for High Performance
  Computing, Networking, Storage and Analysis}, pages 1--14. IEEE, 2020.

\bibitem{han2017ese}
Song Han, Junlong Kang, Huizi Mao, Yiming Hu, Xin Li, Yubin Li, Dongliang Xie,
  Hong Luo, Song Yao, Yu~Wang, et~al.
\newblock Ese: Efficient speech recognition engine with sparse lstm on fpga.
\newblock In {\em Proceedings of the 2017 ACM/SIGDA International Symposium on
  Field-Programmable Gate Arrays}, pages 75--84, 2017.

\bibitem{MAE}
Kaiming He, Xinlei Chen, Saining Xie, Yanghao Li, Piotr Doll\'ar, and Ross
  Girshick.
\newblock Masked autoencoders are scalable vision learners.
\newblock In {\em IEEE Conference on Computer Vision and Pattern Recognition
  (CVPR)}, pages 16000--16009, June 2022.

\bibitem{resnet}
Kaiming He, Xiangyu Zhang, Shaoqing Ren, and Jian Sun.
\newblock Deep residual learning for image recognition.
\newblock In {\em Proceedings of the IEEE conference on computer vision and
  pattern recognition}, pages 770--778, 2016.

\bibitem{kd}
Geoffrey Hinton, Oriol Vinyals, and Jeff Dean.
\newblock Distilling the knowledge in a neural network.
\newblock {\em arXiv preprint arXiv:1503.02531}, 2015.

\bibitem{mobilenet}
Andrew~G Howard, Menglong Zhu, Bo~Chen, Dmitry Kalenichenko, Weijun Wang,
  Tobias Weyand, Marco Andreetto, and Hartwig Adam.
\newblock Mobilenets: Efficient convolutional neural networks for mobile vision
  applications.
\newblock {\em arXiv preprint arXiv:1704.04861}, 2017.

\bibitem{isola2017image}
Phillip Isola, Jun-Yan Zhu, Tinghui Zhou, and Alexei~A Efros.
\newblock Image-to-image translation with conditional adversarial networks.
\newblock In {\em Proceedings of the IEEE conference on computer vision and
  pattern recognition}, pages 1125--1134, 2017.

\bibitem{karimi2023vidaptive}
Pantea Karimi, Sadjad Fouladi, Vibhaalakshmi Sivaraman, and Mohammad Alizadeh.
\newblock Vidaptive: Efficient and responsive rate control for real-time video
  on variable networks.
\newblock {\em arXiv preprint arXiv:2309.16869}, Sep 2023.

\bibitem{celeba}
Tero Karras, Timo Aila, Samuli Laine, and Jaakko Lehtinen.
\newblock Progressive growing of gans for improved quality, stability, and
  variation.
\newblock {\em arXiv preprint arXiv:1710.10196}, 2017.

\bibitem{ffhq}
Tero Karras, Samuli Laine, and Timo Aila.
\newblock A style-based generator architecture for generative adversarial
  networks.
\newblock In {\em Proceedings of the IEEE/CVF conference on computer vision and
  pattern recognition}, pages 4401--4410, 2019.

\bibitem{kastryulin2022piq}
Sergey Kastryulin, Jamil Zakirov, Denis Prokopenko, and Dmitry~V. Dylov.
\newblock Pytorch image quality: Metrics for image quality assessment, 2022.

\bibitem{srvc}
Mehrdad Khani, Vibhaalakshmi Sivaraman, and Mohammad Alizadeh.
\newblock Efficient video compression via content-adaptive super-resolution.
\newblock In {\em Proceedings of the IEEE/CVF International Conference on
  Computer Vision}, pages 4521--4530, 2021.

\bibitem{kuznetsova2020open}
Alina Kuznetsova, Hassan Rom, Neil Alldrin, Jasper Uijlings, Ivan Krasin, Jordi
  Pont-Tuset, Shahab Kamali, Stefan Popov, Matteo Malloci, Alexander
  Kolesnikov, et~al.
\newblock The open images dataset v4: Unified image classification, object
  detection, and visual relationship detection at scale.
\newblock {\em International Journal of Computer Vision}, 128(7):1956--1981,
  2020.

\bibitem{LeeKKCH22}
Doyup Lee, Chiheon Kim, Saehoon Kim, Minsu Cho, and Wook{-}Shin Han.
\newblock Autoregressive image generation using residual quantization.
\newblock In {\em IEEE Conference on Computer Vision and Pattern Recognition
  (CVPR)}, 2022.

\bibitem{li2022mage}
Tianhong Li, Huiwen Chang, Shlok~Kumar Mishra, Han Zhang, Dina Katabi, and
  Dilip Krishnan.
\newblock Mage: Masked generative encoder to unify representation learning and
  image synthesis.
\newblock {\em arXiv preprint arXiv:2211.09117}, 2022.

\bibitem{li2023self}
Tianhong Li, Dina Katabi, and Kaiming He.
\newblock Self-conditioned image generation via generating representations.
\newblock {\em arXiv preprint arXiv:2312.03701}, 2023.

\bibitem{liu2023audioldm}
Haohe Liu, Zehua Chen, Yi~Yuan, Xinhao Mei, Xubo Liu, Danilo Mandic, Wenwu
  Wang, and Mark~D Plumbley.
\newblock Audioldm: Text-to-audio generation with latent diffusion models.
\newblock {\em arXiv preprint arXiv:2301.12503}, 2023.

\bibitem{loshchilov2016sgdr}
Ilya Loshchilov and Frank Hutter.
\newblock Sgdr: Stochastic gradient descent with warm restarts.
\newblock {\em arXiv preprint arXiv:1608.03983}, 2016.

\bibitem{loshchilov2017decoupled}
Ilya Loshchilov and Frank Hutter.
\newblock Decoupled weight decay regularization.
\newblock {\em arXiv preprint arXiv:1711.05101}, 2017.

\bibitem{dvc}
Guo Lu, Wanli Ouyang, Dong Xu, Xiaoyun Zhang, Chunlei Cai, and Zhiyong Gao.
\newblock {DVC: An end-to-end deep video compression framework}.
\newblock In {\em Proceedings of the IEEE Conference on Computer Vision and
  Pattern Recognition}, pages 11006--11015, 2019.

\bibitem{mackay2005fountain}
David~JC MacKay.
\newblock Fountain codes.
\newblock {\em IEE Proceedings-Communications}, 152(6):1062--1068, 2005.

\bibitem{vp9}
Debargha Mukherjee, Jingning Han, Jim Bankoski, Ronald Bultje, Adrian Grange,
  John Koleszar, Paul Wilkins, and Yaowu Xu.
\newblock A technical overview of {VP9}, the latest open-source video codec.
\newblock {\em SMPTE Motion Imaging Journal}, 124(1):44--54, 2015.

\bibitem{netravali2015mahimahi}
Ravi Netravali, Anirudh Sivaraman, Somak Das, Ameesh Goyal, Keith Winstein,
  James Mickens, and Hari Balakrishnan.
\newblock Mahimahi: accurate $\{$Record-and-Replay$\}$ for $\{$HTTP$\}$.
\newblock In {\em 2015 USENIX Annual Technical Conference (USENIX ATC 15)},
  pages 417--429, 2015.

\bibitem{rao2021dynamicvit}
Yongming Rao, Wenliang Zhao, Benlin Liu, Jiwen Lu, Jie Zhou, and Cho-Jui Hsieh.
\newblock Dynamicvit: Efficient vision transformers with dynamic token
  sparsification.
\newblock {\em Advances in neural information processing systems},
  34:13937--13949, 2021.

\bibitem{razavi2019generating}
Ali Razavi, Aaron Van~den Oord, and Oriol Vinyals.
\newblock Generating diverse high-fidelity images with vq-vae-2.
\newblock {\em Advances in neural information processing systems}, 32, 2019.

\bibitem{reed1960polynomial}
Irving~S Reed and Gustave Solomon.
\newblock Polynomial codes over certain finite fields.
\newblock {\em Journal of the society for industrial and applied mathematics},
  8(2):300--304, 1960.

\bibitem{rombach2022high}
Robin Rombach, Andreas Blattmann, Dominik Lorenz, Patrick Esser, and Bj{\"o}rn
  Ommer.
\newblock High-resolution image synthesis with latent diffusion models.
\newblock In {\em Proceedings of the IEEE/CVF Conference on Computer Vision and
  Pattern Recognition}, pages 10684--10695, 2022.

\bibitem{tambur}
Michael Rudow, Francis~Y Yan, Abhishek Kumar, Ganesh Ananthanarayanan, Martin
  Ellis, and KV~Rashmi.
\newblock Tambur: Efficient loss recovery for videoconferencing via streaming
  codes.
\newblock In {\em 20th USENIX Symposium on Networked Systems Design and
  Implementation (NSDI 23)}, pages 953--971, 2023.

\bibitem{h264}
Heiko Schwarz, Detlev Marpe, and Thomas Wiegand.
\newblock Overview of the scalable video coding extension of the h. 264/avc
  standard.
\newblock {\em IEEE Transactions on circuits and systems for video technology},
  17(9):1103--1120, 2007.

\bibitem{sivaraman2022gemino}
Vibhaalakshmi Sivaraman, Pantea Karimi, Vedantha Venkatapathy, Mehrdad Khani,
  Sadjad Fouladi, Mohammad Alizadeh, Fr{\'e}do Durand, and Vivienne Sze.
\newblock Gemino: Practical and robust neural compression for video
  conferencing.
\newblock {\em arXiv preprint arXiv:2209.10507}, 2022.

\bibitem{h265}
Gary~J Sullivan, Jens-Rainer Ohm, Woo-Jin Han, and Thomas Wiegand.
\newblock {Overview of the high efficiency video coding (HEVC) standard}.
\newblock {\em IEEE Transactions on circuits and systems for video technology},
  22(12):1649--1668, 2012.

\bibitem{szegedy2016rethinking}
Christian Szegedy, Vincent Vanhoucke, Sergey Ioffe, Jon Shlens, and Zbigniew
  Wojna.
\newblock Rethinking the inception architecture for computer vision.
\newblock In {\em Proceedings of the IEEE conference on computer vision and
  pattern recognition}, pages 2818--2826, 2016.

\bibitem{efficientnet}
Mingxing Tan and Quoc Le.
\newblock Efficientnet: Rethinking model scaling for convolutional neural
  networks.
\newblock In {\em International conference on machine learning}, pages
  6105--6114. PMLR, 2019.

\bibitem{itu}
International~Telecommunication Union.
\newblock {ITU-T G.1010: End-user multimedia QoS categories}.
\newblock In {\em Series G: Transmission Systems and Media, Digital Systems and
  Networks}, 2001.

\bibitem{OordVK17}
A{\"{a}}ron van~den Oord, Oriol Vinyals, and Koray Kavukcuoglu.
\newblock Neural discrete representation learning.
\newblock In {\em Advances in Neural Information Processing Systems (NeurIPS)},
  2017.

\bibitem{vaswani2017attention}
Ashish Vaswani, Noam Shazeer, Niki Parmar, Jakob Uszkoreit, Llion Jones,
  Aidan~N Gomez, {\L}ukasz Kaiser, and Illia Polosukhin.
\newblock Attention is all you need.
\newblock {\em Advances in neural information processing systems}, 30, 2017.

\bibitem{Maxine}
Ting-Chun Wang, Arun Mallya, and Ming-Yu Liu.
\newblock One-shot free-view neural talking-head synthesis for video
  conferencing.
\newblock In {\em Proceedings of the IEEE/CVF Conference on Computer Vision and
  Pattern Recognition}, pages 10039--10049, 2021.

\bibitem{talkinghead}
Ting-Chun Wang, Arun Mallya, and Ming-Yu Liu.
\newblock One-shot free-view neural talking-head synthesis for video
  conferencing.
\newblock In {\em Proceedings of the IEEE/CVF conference on computer vision and
  pattern recognition}, pages 10039--10049, 2021.

\bibitem{ssim}
Zhou Wang, Alan~C Bovik, Hamid~R Sheikh, and Eero~P Simoncelli.
\newblock Image quality assessment: from error visibility to structural
  similarity.
\newblock {\em IEEE transactions on image processing}, 13(4):600--612, 2004.

\bibitem{nas}
Hyunho Yeo, Youngmok Jung, Jaehong Kim, Jinwoo Shin, and Dongsu Han.
\newblock Neural adaptive content-aware internet video delivery.
\newblock In {\em 13th {USENIX} Symposium on Operating Systems Design and
  Implementation ({OSDI} 18)}, pages 645--661, 2018.

\bibitem{yu2021vector}
Jiahui Yu, Xin Li, Jing~Yu Koh, Han Zhang, Ruoming Pang, James Qin, Alexander
  Ku, Yuanzhong Xu, Jason Baldridge, and Yonghui Wu.
\newblock Vector-quantized image modeling with improved vqgan.
\newblock {\em arXiv preprint arXiv:2110.04627}, 2021.

\bibitem{lpips}
Richard Zhang, Phillip Isola, Alexei~A Efros, Eli Shechtman, and Oliver Wang.
\newblock The unreasonable effectiveness of deep features as a perceptual
  metric.
\newblock In {\em Proceedings of the IEEE conference on computer vision and
  pattern recognition}, pages 586--595, 2018.

\bibitem{zhang2020sparch}
Zhekai Zhang, Hanrui Wang, Song Han, and William~J Dally.
\newblock Sparch: Efficient architecture for sparse matrix multiplication.
\newblock In {\em 2020 IEEE International Symposium on High Performance
  Computer Architecture (HPCA)}, pages 261--274. IEEE, 2020.

\bibitem{zheng2001improved}
Haitao Zheng and Jill Boyce.
\newblock An improved udp protocol for video transmission over
  internet-to-wireless networks.
\newblock {\em IEEE Transactions on Multimedia}, 3(3):356--365, 2001.

\end{thebibliography}

\clearpage

\appendix
\section{Implementation Details}
\label{sec:implementation}

Because of GPU memory limitation, we adopt a two-stage training recipe for \namereparo similar to many prior approaches~\cite{li2022mage, esser2021taming}. We first train our VQGAN codec which encodes each video frame into discrete tokens \emph{without} any losses. We then \emph{fix} the VQGAN codec and train the loss recovery module on the discrete tokens with self-dropping and packet loss. In this section, we describe the neural network structure and training schemes of \namereparo's neural codec and loss recovery module, as well as the design of the bitrate controller in detail.

\subsection{Neural Codec}
\label{sec:codec-implementation}

\begin{figure}[h]
\begin{center}
\includegraphics[width=1.0\linewidth]{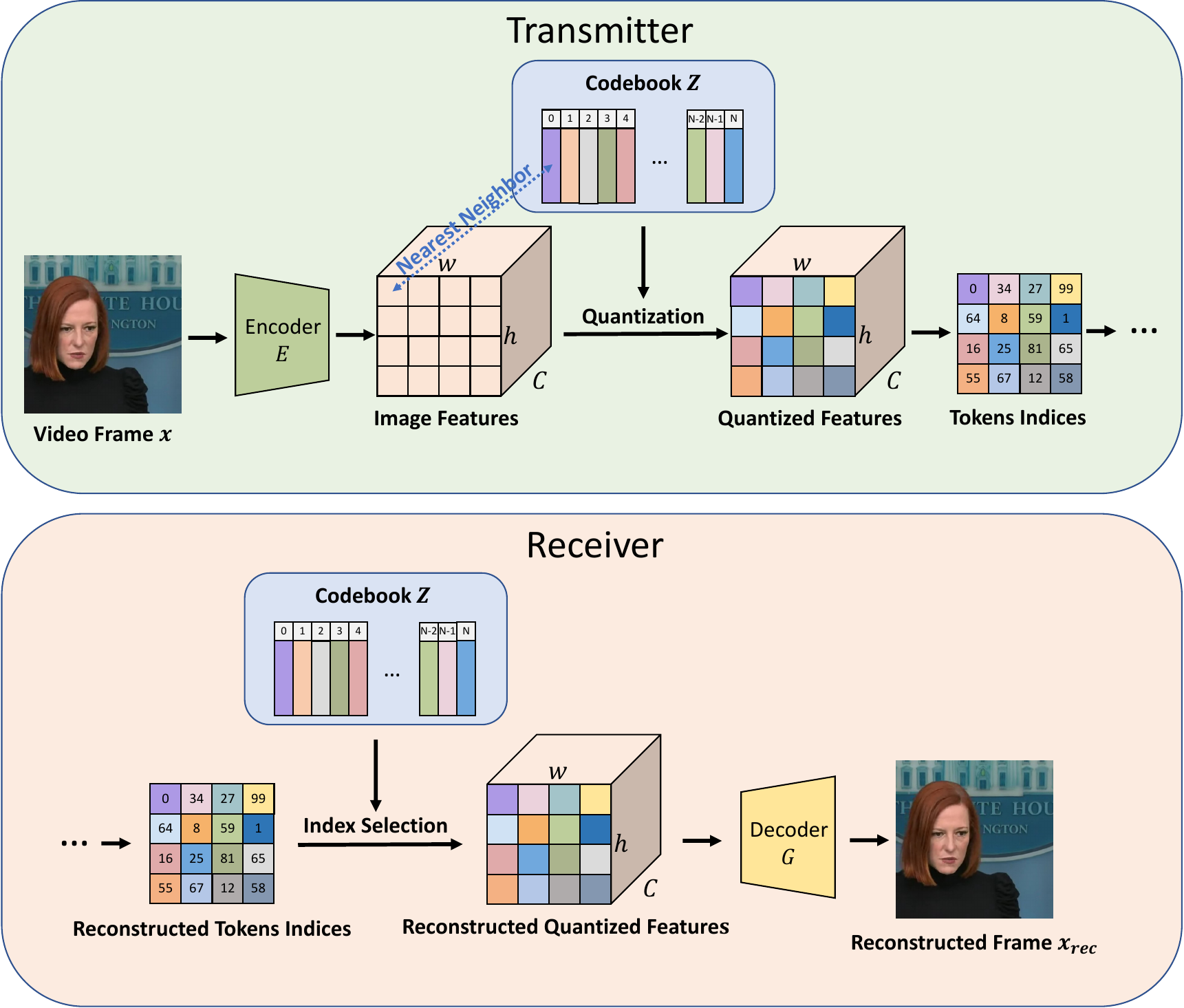}
\end{center}
    \caption{Token-based neural codec. The transmitter first uses the encoder to convert patches from video frames into features. It then uses a codebook to quantize the features into tokens by finding the nearest neighbor of each feature in the codebook. The receiver first maps the received and generated tokens to back-to-image features using the codebook. It then uses a decoder to reconstruct the video frame.}
    \label{fig:vqgan-appendix}
\end{figure}

\noindent\textbf{Model Structure.} We use a CNN-based VQGAN \cite{esser2021taming} encoder and quantizer to tokenize the $3\times512\times512$ input frame to $128\times32\times32$ quantized features, where 128 is the number of channels of the quantized features. It then uses a codebook to quantize the features by finding the nearest neighbor of each feature in the codebook. The codebook is a $1024\times 128$ matrix by default, with 1024 entries, each of which uses a 128-dimensional feature. The decoder operates on the quantized features and reconstructs the $3\times512\times512$ video frame. The encoder consists of 5 blocks and each block consists of 2 residual blocks which follow standard ResNet's residual block design \cite{resnet}. After each block in the encoder, the feature vector is down-sampled by 2 using average pooling. The quantizer then maps each pixel of the encoder's output feature map to the nearest token (based on $L_2$ distance) in the codebook $Z$ with $N=1024$ entries, each entry with $128$ channels. The decoder consists of another 5 blocks where each encoder block has 2 residual blocks. After each block in the decoder, the feature map is up-sampled by 2 using bicubic interpolation. The tokenizer consists of 23.8M parameters and the detokenizer consists of 30.5M parameters.

\noindent\textbf{Training schemes.} We follow the original VQGAN training recipe \cite{esser2021taming} to train the VQGAN. We use a vector-quantize loss between the image features and quantized tokens that nudges the image features towards the tokens that they map to, a reconstruction loss ($L_1$) between the input and final reconstructed frame, a perceptual loss~\cite{lpips} between the input and reconstructed frame, and a discriminative loss \cite{isola2017image} between the input and reconstructed frame. Detailed descriptions of the losses can be found in the VQGAN paper \cite{esser2021taming}.

We use the officially released VQGAN encoder and decoder pre-trained on OpenImages \cite{kuznetsova2020open} to initialize our codec whenever possible. OpenImages is a large-scale image dataset consisting of $\sim$9M natural images. We observe that such initialization largely speeds up our training (takes $\sim$ 10 epochs to converge), but we also note that training from scratch on our pre-training face datasets can achieve similar performance with a much longer training time ($\sim$200 epochs). We train our neural codec using a constant learning rate and train it until there is no substantial change in the training loss. Please refer to \Tab{codec-setting} for the training recipe of our VQGAN codec.

\begin{table}[t]
\caption{\textbf{VQGAN codec training setting.}}
\label{tab:codec-setting}
\begin{center}{
\small
\begin{tabular}{l|l}

Parameter & Value \\
\toprule

Optimizer & Adam \cite{loshchilov2017decoupled} \\
Base Learning Rate & 1e-4 \\ 
Weight Decay & 0 \\
Optimizer Momentum & $\beta_1, \beta_2=0.5, 0.9$ \\
Batch Size & 24 \\
Learning Rate Schedule & Constant \cite{loshchilov2016sgdr} \\
Warmup Epochs & 0 \\
Gradient Clip & 0 \\
Dropout & 0 \\

\end{tabular}
}
\end{center}
\vspace{-10pt}
\end{table}

\subsection{Loss Recovery Module}
\label{sec:loss-recovery-implementation}

\noindent\textbf{Model Structure.} The major component of our loss recovery module is  a spatio-temporal ViT network. In our default setting, the input tokens are of shape $C\times T \times h \times w$, where $C=768$, $T=6$, $h=32$, $w=32$. We use two separate learnable position embeddings, one for time and one for space, which we add together to provide each input token its positional information. 
We then adopt a standard spatio-temporal ViT architecture \cite{vit}, which consists of a stack of spatio-temporal Transformer blocks \cite{vaswani2017attention}. Each spatio-temporal block consists of a spatial block and a temporal block. Each of the two blocks independently consists of a multi-head self-attention block and a multi-layer perceptron (MLP) block. In total, we use 20 spatio-temporal Transformer blocks. The number of heads in each multi-head self-attention layer is 12, and the MLP ratio is 4. The embedding dimension throughout the Transformer is 768. Our spatio-temporal ViT consists of 172M parameters. We note that more Transformer blocks, more heads in the self-attention layer, and a larger embedding dimension can further improve the performance of \namereparo~, but they also introduce more computation overheads.

\noindent\textbf{Training schemes.} \Tab{pretrain-setting} provides the training recipe for our spatio-temporal ViT for loss recovery. The self-drop rate is sampled from a truncated Gaussian distribution from 0 to 0.6 and centered at 0.3, with a standard deviation of 0.3. The packet loss rate is uniformly sampled from 0 to 0.8.

\begin{table}[t]
\caption{\textbf{Loss recovery module training setting.}}
\label{tab:pretrain-setting}
\begin{center}{
\small
\begin{tabular}{l|l}

Parameter & Value \\
\toprule

Optimizer & Adam \cite{loshchilov2017decoupled} \\
Learning Rate & 1.5e-5 \\ 
Weight Decay & 0.05 \\
Optimizer Momentum & $\beta_1, \beta_2=0.9, 0.95$ \\
Batch Size & 24 \\
Learning Rate Schedule & Cosine Decay \cite{loshchilov2016sgdr} \\
Warmup Epochs & 10 \\
Training Epochs & 200 \\
Gradient Clip & 3.0 \\
Label Smoothing \cite{szegedy2016rethinking} & 0.1 \\
Dropout & 0.1 \\
Min. Self-Drop Rate & 0 \\
Max. Self-Drop Rate & 0.6 \\
Self-Drop Rate Mode & 0.3 \\
Self-Drop Rate Std. Dev. & 0.3 \\
Min. Packet Loss Rate & 0 \\
Max. Packet Loss Rate & 0.8 \\

\end{tabular}
}
\end{center}
\vspace{-10pt}
\end{table}

\subsection{Bitrate Controller}

\namereparo employs self-dropping to drop a fixed fraction of tokens across all packets of a frame to achieve the target bitrate. For example, if the target bitrate is \SI{200}{Kbps} and the bitrate when transmitting all tokens is \SI{300}{Kbps}, the bitrate controller will sample one-third of the tokens in each packet to drop.

To minimize the impact of self-dropping on the loss recovery module, we drop tokens randomly in each packet, so that the dropped tokens are distributed uniformly in space and time. However, randomly dropping tokens in each packet requires telling the receiver which tokens are dropped, leading to bandwidth overheads. Otherwise, the receiver will be confused about the position of each received token in the $h\times w$ token map.

To address this issue, we deterministically sample the tokens to be self-dropped in each packet based on the frame index and packet index. We achieve this by setting the random seed for pseudo-random self-dropping sampling in a packet to $4\times\text{frame index}+\text{packet index}$. Consequently, the receiver compares the received packet size to the expected packet size to identify how many tokens were lost. The receiver then decodes the locations of the lost tokens by simulating the self-drop procedure on its end by repeating the pseudo-random sampling procedure with the same seed and drop rate as the transmitter.

At the start of a video conference, the transmitter selects the variant of the codec and loss recovery module to use based on the target bitrate and synchronizes this information with the receiver. It also communicates the expected number of tokens per packet and frame during this process. Once the variant is established, it can adapt to bitrate changes of up to 50\% with self-dropping. If the target bitrate changes significantly, the transmitter selects a new variant and notifies the receiver.

\end{document}